\documentclass[acmsmall]{acmart}

\AtBeginDocument{%
  \providecommand\BibTeX{{%
    \normalfont B\kern-0.5em{\scshape i\kern-0.25em b}\kern-0.8em\TeX}}}

\setcopyright{acmcopyright}
\copyrightyear{2024}
\acmYear{2024}
\acmDOI{XXXXXXX.XXXXXXX}

\acmJournal{TOIS}
\acmVolume{00}
\acmNumber{0}
\acmArticle{001}
\acmMonth{0}

\usepackage{algorithm} 
\usepackage{algorithmic}
\usepackage{booktabs}
\usepackage{clrscode3e}
\usepackage{hyperref}
\usepackage{multirow}
\usepackage{graphicx}
\usepackage{subfigure}
\usepackage{textcomp}
\usepackage{xcolor}
\usepackage{acronym}
\usepackage{amsmath}

\usepackage{amssymb}
\usepackage{amsfonts}
\usepackage{amsthm}
\usepackage{mathrsfs}
\usepackage{indentfirst}
\usepackage{cases}
\usepackage{bm}
\usepackage{color}
\usepackage{stfloats}
\usepackage{url}
\usepackage{verbatim}
\usepackage{array}

\acrodef{SR}{Sequential Recommendation}
\acrodef{SRs}{Sequential Recommender systems}
\acrodef{LR}{Lightweight Recommendation}
\acrodef{GSRs}{Graph-based Sequential Recommender systems}
\acrodef{RNN}{Recurrent Neural Network}
\acrodef{GRU}{Gated Recurrent Unit}
\acrodef{MLP}{Multi-Layer Perceptron}
\acrodef{DNN}{Deep Neural Network}
\acrodef{MF}{Matrix Factorization}
\acrodef{GCN}{Graph Convolutional Network}
\acrodef{GNN}{Graph Neural Network}
\acrodef{EA}{External Attention}
\acrodef{GAN}{Generative Adversarial Network}
\acrodef{LLMs}{Large Language Models}
\acrodef{CV}{Computer Vision}
\begin{document}


\title{Lightweight yet Efficient: An External Attentive Graph Convolutional Network with Positional Prompts for Sequential Recommendation}
\author{Jinyu Zhang}
\email{jinyuz1996@outlook.com}
\affiliation{%
  \institution{College of Computer Science and Engineering, Shandong University of Science and Technology}
  \city{Qingdao}
  \country{China}
  \postcode{266590}
}

\author{Chao Li}
\email{lichao@sdust.edu.cn}
\affiliation{%
  \institution{College of Computer Science and Engineering, Shandong University of Science and Technology}
  \city{Qingdao}
  \country{China}
  \postcode{266590}
}

\author{Zhongying Zhao}
\authornote{Corresponding Author: Zhongying Zhao.}
\email{zzysuin@163.com, zyzhao@sdust.edu.cn}
\affiliation{%
  \institution{College of Computer Science and Engineering, Shandong University of Science and Technology}
  \city{Qingdao}
  \country{China}
  \postcode{266590}
}


\begin{abstract}
Graph-based Sequential Recommender systems (GSRs) have gained significant research attention due to their ability to simultaneously handle user-item interactions and sequential relationships between items. Current GSRs often utilize composite or in-depth structures for graph encoding (e.g., the Graph Transformer). Nevertheless, they have high computational complexity, hindering the deployment on resource-constrained edge devices. Moreover, the relative position encoding in Graph Transformer has difficulty in considering the complicated positional dependencies within sequence. To this end, we propose an \underline{E}xternal \underline{A}ttentive \underline{G}raph convolution network with \underline{P}ositional prompts for \underline{S}equential recommendation, namely EA-GPS.
Specifically, we first introduce an external attentive graph convolutional network that linearly measures the global associations among nodes via two external memory units. Then, we present a positional prompt-based decoder that explicitly treats the absolute item positions as external prompts. By introducing length-adaptive sequential masking and a soft attention network, such a decoder facilitates the model to capture the long-term positional dependencies and contextual relationships within sequences. Extensive experimental results on five real-world datasets demonstrate that the proposed EA-GPS outperforms the state-of-the-art methods. Remarkably, it achieves the superior performance while maintaining a smaller parameter size and lower training overhead. The implementation of this work is publicly available at \url{https://github.com/ZZY-GraphMiningLab/EA-GPS}.
\end{abstract}

\ccsdesc[500]{Information systems~Recommender systems}
\ccsdesc[500]{Information systems~Collaborative filtering}
\ccsdesc[500]{Information systems~Personalization}

\keywords{Sequential Recommendation, Graph Convolutional Network, Lightweight Recommendation, Prompt-based Learning, External Attention}

\received{18 January 2024}
\received[revised]{XX XXXX 2024}
\received[accepted]{XX XXXX 2024}

\maketitle

\section{Introduction}\label{sec:introduction}
\noindent With the widespread application of online shopping \cite{ZhangPrompt2023}, information retrieval \cite{WangMCZLM23}, and mobile services \cite{lea23dasfaa}, a vast amount of data about user behaviors is now routinely recorded by platforms and servers. As we all know, user behaviors are typically not fragmented or independent; instead, they manifest as historical sequences via temporal accumulation \cite{hidasi2016srnn}. 
In light of this, scholars develop Sequential Recommender systems (SRs) to forecast user consumption based on their sequential preferences \cite{seq23wwwmlp}. By modeling the evolution of user behaviors, SRs have the capability to provide more accurate and relevant recommendations. The rationale behind SR is that user preferences and behaviors are not static but evolving \cite{seq22icdmrnn}.
Therefore, by analyzing the temporal patterns and trends in user behaviors, SR can offer recommendations that are more in line with the user’s current interests and needs.
It has been validated by extensive research, demonstrating the significance of simultaneously considering collaborative filtering signals and sequential preferences. \cite{ShiLWZ23,YangHXHLL23,DuYZF0LS023,ZhangWC023}.

\begin{figure}[b]
\centering
\includegraphics[width=0.7\columnwidth]{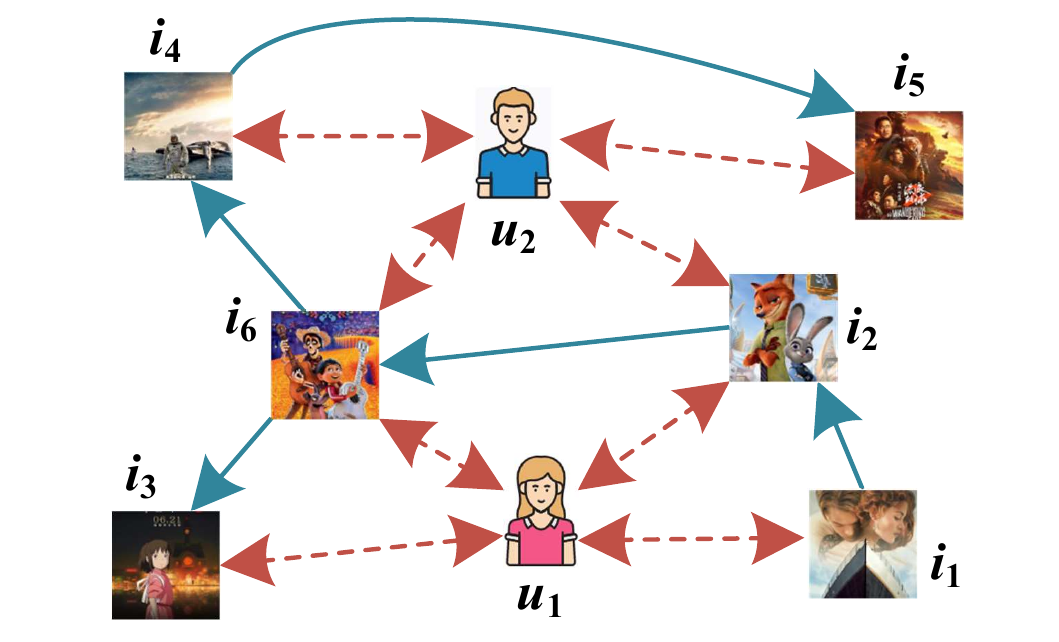} 
\caption{An example of the sequential graph, where user-item interactions are denoted by red arrows and sequential dependencies between items are in blue.}
\label{fig:illustration}
\Description{An example of a sequential graph showing user-item interactions in red and sequential dependencies in blue.}
\end{figure}

In the early exploration of \ac{SR} tasks, Recurrent Neural Network (RNN)-based models are frequently employed to measure the sequential dependencies in user historical behaviors \cite{seq22icdmrnn}. However, RNNs have difficulties in considering the long-term associations among user interactions. 
Then, researchers sought to enhance Sequential Recommender systems (SRs) by employing Generative Adversarial Networks (GANs) \cite{seq21dasfaagan} and Transformers \cite{ZhangWC023, SunLWPLOJ19}. These methods are designed to capture the evolution of users’ sequential patterns more effectively. Despite these advancements, their performance in extracting structural information from sequential transitions is still limited. 
To address the above issues, some researchers attempt to study \ac{SR} via graph-based techniques \cite{HsuL21,xia_self_AAAI_2021}, a.k.a. \ac{GSRs}. GSRs leverage the power of graph structures to aggregate complex associations between users and items, encompassing user-item interactions and sequential relationships (as depicted in Fig. \ref{fig:illustration}). Nevertheless, GSRs are still limited when considering the global correlations among items. Hence, scholars attempt to combine the Transformer with Graph Neural Networks (GNNs) to create the Graph Transformer \cite{XiaHXP23, LiXRY0023}. The self-attention mechanism in the Transformer enables global weighting for capturing long-range dependencies in sequential graphs. Although Transformer-based GSRs have achieved improved performance on SR tasks, they still face the following challenges:

\textbf{(1) The global weighting strategy of self-attention mechanism requires high computational complexity.} 
Most Transformer-based \ac{GSRs} uniformly employ graph encoders with self-attention mechanism to measure the global associations among nodes. The core idea of such methods is to consider the overall contextual associations by calculating the attention weight between pairwise nodes in sequential graphs \cite{LiXRY0023}. Therefore, the \ac{GSRs} are indeed good at handling long-distance dependencies and capturing global information. However, they typically demand quadratic computational complexities (i.e., $\mathcal{O}(N^2 \times d)$), posing a significant challenge for model deployment on resource-constrained edge devices \cite{LiCZY21}. 

\textbf{(2) The relative position encoding has difficulty in considering the complicated positional dependencies.}
Transformer-based GSRs with relative position encoding have been proven effective in modeling sequences with varying lengths. The core idea of relative position encoding is to calculate the relative positions between adjacent items in the sequence \cite{HuangMLDWL23}. This strategy is based on the "locality of position" assumption, i.e., an item has a direct dependency relationship with its adjacent items \cite{ShawUV18}. However, the expression of user preferences within a sequence is sometimes discontinuous or periodic. Relative positional encoding merely focuses on the continuous positional correlations, overlooking the long-term or intermittent positional dependencies of items within a sequence \cite{WuPCFC21}. Such an overlooking makes it challenging for GSRs to comprehend the positional relationships of items throughout the entire sequence.

To this end, we propose an \textbf{E}xternal \textbf{A}ttentive \textbf{G}raph convolutional network with \textbf{P}ositional prompts for \textbf{S}equential recommendation, namely EA-GPS. 
Specifically, we first introduce an external attentive graph convolutional network. It contains an External Attention (EA)-based external encoder, which operates in parallel with the primary graph encoder. Such an external encoder enables a linear measurement of global associations among nodes via two external memory units. The independent yet reusable external memory units also facilitate sharing global attention weights across convolution layers.
Then, we propose a positional prompt-based decoder to consider the complicated positional dependencies of interactions while learning sequence representations. This decoder first treats the absolute item positions as prompts. Subsequently, it operates a vertical concatenation between sequence and prompt representations, which directly tells the model about the exact position of items within sequences. By employing a length-adaptive sequential masking and a soft attention network, the decoder could better understand the long-term positional dependencies and contextual relationships of items.

The main contributions of this work can be briefly described as:
\begin{itemize}
    \item We investigate SR in an emerging yet challenging scenario (i.e., the lightweight graph-based sequential recommendation scenario). After highlighting the challenges faced by existing GSRs, we propose a graph-based solution called EA-GPS.
    \item We devise a lightweight external encoder to linearly measure the global associations among nodes. This design relieves the primary graph encoder from high computational complexity.
    \item We present an efficient prompt-based sequence decoder that treats the exact position of items as prompts to facilitate capturing the complicated positional dependencies and contextual relationships of items.
    \item We conduct extensive experiments on five real-world datasets. The experimental results fully demonstrate that EA-GPS requires a smaller scale of parameters and lower training consumption, but outperforms the state-of-the-art GSRs.
\end{itemize}

\section{Related Work}
\noindent This section considers four types of recommendation methods, i.e., Sequential Recommendation, Graph-based Sequential Recommendation, Prompting-based Recommendation, and Lightweight Recommendation.

\subsection{Sequential Recommendation}
\noindent \acf{SRs} aim to capture the evolution of users' behavioral preferences by modeling their interaction sequences \cite{seq23wwwmlp}. Early explorations on Sequential Recommendation (SR) mainly focus on \ac{RNN}-based \cite{hidasi2016srnn,quadrana2017hrnn,seq22icdmrnn} structures. For example, Hidasi et al. \cite{hidasi2016srnn} are the first group that exploits RNN to measure user preferences in dynamic sequences. Quadrana et al. \cite{quadrana2017hrnn} propose a hierarchical structure to relay and evolve the hidden states, which achieves better performance on \ac{SR} tasks. However, these works meet gradient vanishing problems while considering the item associations from long sequences \cite{HuangW0023}.
A practical solution is to leverage Generative Adversarial Networks (GANs) \cite{seq21dasfaagan} to maintain long-term memory and overcome the gradient block problem. GANs are good at learning complex data distributions, enabling the creation of high-quality synthetic data for sequential recommendation tasks \cite{NiZWHQ23}. However, a significant drawback of GANs is their training instability, which can lead to mode collapse and difficulty in convergence. Then, some scholars consider utilizing contrastive learning \cite{YangHXHLL23, WangMCZLM23} to improve the quality of long-term sequential preferences. By contrasting similar and dissimilar instances, this strategy encourages the model to capture the underlying structure and relationships within sequences \cite{TianHZ0Z23}. Another popular solution is the Transformer \cite{LaiCY0C023,ZhangWC023}, which significantly improves the accuracy and robustness of sequential recommendation via the self-attention mechanism. Such a component allows the Transformer to capture long-range dependencies in sequences, enabling better performance on tasks requiring context understanding \cite{HeHSLJC18}. These deep learning-based networks achieve great success in \ac{SR} tasks but are limited in excavating structural information inside the sequential transitions.

\subsection{Graph-based Sequential Recommendation}
\noindent As one of the most popular solutions for Sequential Recommendation (SR) tasks, Graph-based Sequential Recommender systems (GSRs) have demonstrated efficacy in extracting structural information from complicated user-item interactions by constructing sequential graphs \cite{ZhangLXXLHCM22, ChangGZHNSJ021}. For example, PTGCN \cite{HuangMLDWL23} captures sequential patterns and temporal dynamics of user-item interactions by incorporating a position-enhanced and time-aware graph convolutional operator. It simultaneously learns the dynamic representations of users and items on a bipartite graph, utilizing a self-attention aggregator. RetaGNN \cite{HsuL21} is a relational attentive GNN that operates on local subgraphs derived from user-item pairs. It distinguishes itself by allocating learnable weight matrices to the various relationships between users, items, and attributes, instead of applying them to nodes or edges.
However, these models are limited when handling large-scale graphs. 
To address the above limitation, researchers start using more complicated structures (such as Hypergraphs \cite{xia_self_AAAI_2021} or Graph Contrastive Learning \cite{WuWF0CLX21}).
For example, Wu et al. \cite{WuWF0CLX21} propose a contrastive learning-based method that supplements the traditional sequential recommendation tasks with an auxiliary self-supervised task, reinforcing node representation learning via self-discrimination.
Xia et al. \cite{xia_self_AAAI_2021} integrate hypergraph and contrastive learning to capture the beyond-pairwise relations on sequential graphs. By exploiting a dual-channel hypergraph convolution network, it successfully models the complex high-order information among items. 
These GSRs achieve great success in SR tasks but have difficulty capturing global correlations among nodes from sequential graphs. To this end, scholars have started incorporating Transformer with GCNs \cite{XiaHXP23, LiXRY0023}. For example, Xia et al. \cite{XiaHXP23} introduce a temporal graph transformer to concurrently capture varying short-term and long-range user-item interaction patterns on sequential graphs. Nevertheless, the global weighting component (i.e., the self-attention mechanism) in Graph Transformer has quadratic computational complexity, hindering the deployment of GSRs on resource-constrained edge devices \cite{lea23dasfaa}.

\subsection{Lightweight Recommendation}
\noindent Lightweight recommendation algorithms aim to provide efficient and accurate recommendations by leveraging various techniques to reduce computational complexity and storage requirements, making them suitable for real-world applications with constrained resources \cite{XvLL0LH22,ZhouYXBW23,KnyazevO23}. There are some common strategies, including matrix factorization techniques like Singular Value Decomposition (SVD) \cite{kalman1996singularly} and neighborhood-based algorithms like k-nearest neighbors (KNN). These shallow traditional machine learning methods have been proven effective in the early exploration of dimensionality reduction. However, with the advancement of deep neural networks and the increasing complexity of problems, traditional lightweight methods have become less adaptive. Nonetheless, these concepts continue to guide current explorations \cite{ShiLWZ23}. Existing studies on \ac{LR} can be classified into two categories: 
1) The one aims to reduce computational complexity. By refining algorithms or simplifying model structures, this strategy directly alleviates the scale of parameters \cite{lea23dasfaa,ZhouYXBW23}. 
For example, Zhou et al. \cite{ZhouYXBW23} proposes a novel lightweight matrix factorization for recommendations that deploys shared gradients training on local networks, serving as a two-phase solution to protect the security of users’ data and reduce the dimension of the items. 
Lian et al. \cite{LianWLLC020} provide a novel solution to refine the backbone network, which employs an indirect coding approach. It reduces the computational cost of representation learning by maintaining a coding dictionary.
2) The other aims to remove unnecessary components \cite{MeiZK22}. This strategy can also effectively release the parameter scale but requires verifying the importance of components \cite{YanLZW22,KnyazevO23,MiaoLY22}. For example, Yan et al. \cite{YanLZW22} exploit bidirectional bijection relation-type modeling to enable scalability for large graphs. This method removes the constraints of negative sampling on knowledge graphs, simplifying the computational complexity. 
Miao et al. \cite{MiaoLY22} remove the transmission structure between social and interactive graphs in traditional social recommendation tasks. They fuse the social relationships and interactions into a unified heterogeneous graph to encode high-order collaborative signals explicitly, significantly reducing the computational complexity.  
Although lightweight concepts become popular in traditional recommendation tasks, research on lightweight sequential recommender systems (especially on GSRs) remains largely unexplored \cite{LiCZY21}.

\subsection{Prompt-based Recommendation}
\noindent Prompt-tuning paradigm is initially proposed in Natural Language Processing (NLP), which adapts the Pre-trained Language Models (PLMs) to the specific downstream tasks (especially in few-shot scenarios) \cite{LiuJFTDY022,BrownMRSKDNSSAA20}.
One research direction is to use rigid prompting (token-level) templates, which involve manually designing prompts and splicing them into token sequences \cite{GuHLH22}. In the domain of recommendation systems, numerous studies have employed rigid templates for the application of prompt-based learning. The methods based on rigid prompting templates usually convert the input features to natural language sequences \cite{Geng0FGZ22, ShinRLWS20}. Then, they construct the rigid templates by concatenating the input features with language prompts (e.g., descriptions of the downstream tasks). By recasting recommendations as cloze-style mask-prediction tasks, these methods could stimulate the potentials of large pre-trained models, thereby achieving better performance on recommendation tasks \cite{ZhangW23}.
In contrast, soft prompting (embedding-level) templates consist of randomly initialized learnable continuous embeddings, frequently adopted in recent studies \cite{WuXZZZLH24,BrownMRSKDNSSAA20, ZhangPrompt2023}. For instance, Wu et al. \cite{WuXZZZLH24} build personalized soft prompts by mapping user profiles to embeddings and enabling sufficient masked training on prompting templates via prompt-oriented contrastive learning.
As we know, the core idea of prompt-tuning involves creating appropriate prompts that can guide the pretrained model to generate desired predictions \cite{Geng0FGZ22}. Inspired by this idea, scholars are currently delving into the methodology of leveraging “prompts” to augment the representation learning capabilities of traditional deep learning models \cite{dongPrompt2024, Luoprompt2023}. For example, Luo et al. \cite{Luoprompt2023} treat the embeddings of timestamp as external prompts to unearth the potentials of a Transformer-based model. Note that, the purpose of using external prompts is to guide the model to generate outputs that meet specific requirements without altering the model itself \cite{Luoprompt2023} while the objective of prompt-tuning paradigm is to enhance the performance of pre-trained models by optimizing the input prompts, which involves indirect adjustments to the model's parameters \cite{dongPrompt2024}. In this work, we extract the positional information of items within each sequence as external prompts to enhance the sequence representation learning, thereby capturing the complicated positional dependencies of items within sequences.

\begin{table}[b]
\centering
\caption{The notations mainly used in this paper.}
\label{tab:notations}
\footnotesize
\begin{tabular}
{p{2cm}|p{8.5cm}}
\toprule
\midrule
\textbf{Notations} &  \textbf{Descriptions} \\ 
\midrule
$\hat{\bm{E}}^{(l)}$ & The node embeddings on $l$-th layer. \\
$\hat{\bm{E}}_{U}^{(l)}$ & The embeddings of users on $l$-th layer. \\
$\hat{\bm{E}}_{S}^{(l)}$ & The embeddings of items on $l$-th layer. \\
$\bm{E}_{S_k}$ & The embeddings of sequence $S_k$. \\
$\bm{M_k}$ & The external unit acts as the key matrix in attention. \\
$\bm{M_v}$ & The external unit acts as the value matrix in attention.\\
$\bm{A}_{S}$ & The attention map of self-attention.\\
$\bm{A}_{L}$ & The attention map of linear attention.\\
$\bm{A}_{E}$ & The attention map of external attention.\\
$\bm{Z}_{S_k}$ & The embeddings of sequence $S_k$ refined by attention networks. \\
$\bm{\hat{Z}}_{S_k}$ & The embeddings of sequence $S_k$ refined by multi-head external attention. \\
$\bm{\hat{Z}}_{S}$ & The refined embeddings of items across all sequences. \\
$\bm{E}_{S}^{(l)}$ & The updated embeddings of items on $l$-th layer. \\
$\bm{E}_{U}$ & The final node embeddings for users. \\
$\bm{E}_{S}$ & The final node embeddings for items. \\
$\hat{\bm{E}}_{P_k}$ & The initialized positional prompting embeddings for $S_k$. \\
$\bm{E}_{P_k}$ & The prompting embeddings of $S_k$ that has same dimension with $E_{S_k}$. \\
$\bar{\bm{E}}_{X_k}$ & The embeddings of positional prompting template for $S_k$. \\
$\hat{\bm{E}}_{X_k}$ & The template embeddings for $S_k$ with sequential masks. \\
$\bm{E}_{X_k}$ & The sequence embeddings for $S_k$ refined by soft attention network. \\
$\bm{E}_{u_k}$ & The embeddings of user $u_k$ learned by the external attentive graph convolutional network.\\
$\bm{H}_{S_k}$ & The user-specific sequence-level embeddings for user $u_k$. \\
$\bm{H}_{S}$ & The final embeddings for prediction. \\
\midrule
\bottomrule
\end{tabular}
\end{table}

\section{Methodologies\label{sec:methologies}}
\subsection{Preliminaries\label{subsec:preliminaries}}
\noindent 
In this section, we present the notations mainly used in EA-GPS and then define the \ac{SR} tasks. Suppose that $\mathcal{U}=\{u_1, u_2, \dots, u_k, \dots, u_n\}$ is the set of users, where $u_k$ denotes $k$-th individual user $(k=1, 2, 3, \dots, n)$. Similarly, we define $\mathcal{I}=\{I_1, I_2, \dots, I_i,\dots, I_m\}$ as the set of all items, where $I_i$ represents the $i$-th item $(i=1, 2, 3, \dots, m)$.
Let $\mathcal{S}= \{S_1, S_2, \dots, S_k, \dots, S_n\}$ be the interactive sequences of each user, where $S_k= \{v_1, v_2, \dots, v_{t_k}\}$ denotes the historical behavioral sequence of $u_k$ that contains several interactive items $v$ ordered by their timestamps, $t_k \in \mathcal{T}$ denotes the length of $S_k$, $\mathcal{T} = \{t_1, t_2, \dots, t_k, \dots, t_n\}$ denotes the lengths of each sequence. And we also define $P_k=\{p_1, p_2, \dots, p_{t_k}\}$ as the exact position of each interaction in $S_k$, which has been recorded during the data preprocessing.
In addition to the above definitions, the utilized embedded vectors or matrices are presented in Table \ref{tab:notations}.

The purpose of \acf{SR} task is to predict the next item $I_{i+1}$ that user $u_k$ is most likely to click, based on her/his historical interaction sequences $S_k$ \cite{seq23wwwmlp}. The probabilities of all candidate items can be denoted as:
\begin{align}
& P(I_{i+1}|S_k, u_k)\sim f(S_k, u_k),
\end{align}
where $P(I_{i+1}|S_k, u_k)$ denotes the probability of recommending $I_{i+1}$ as the next interested item to user $u_k$ based on her historical interaction sequence $S_k$, $f(S_k, u_k)$ is the function exploited to estimate the probability.
\begin{figure}[t]
\centering
\includegraphics[width=1.0\columnwidth]{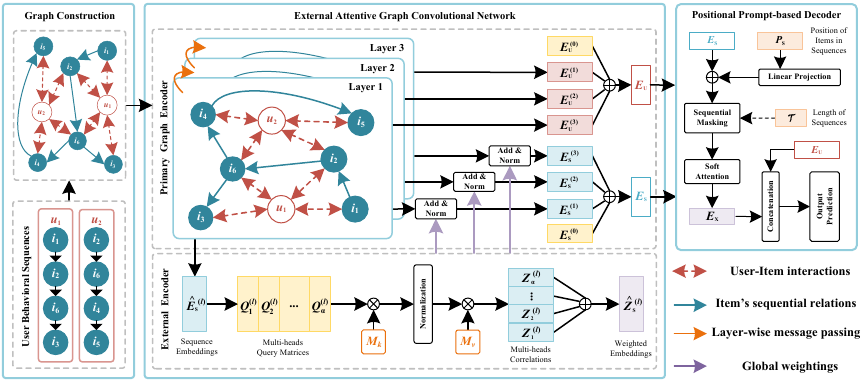}
\caption{Framework of EA-GPS, where $u_1$ and $u_2$ are two different users, and $\{i_1, i_2, \dots, i_6\}$ are the interacted items that compose the behavioral sequences for these users.}
\label{fig:framework}
\Description{Framework of EA-GPS, where $u_1$ and $u_2$ are two different users, and $\{i_1, i_2, \dots, i_6\}$ are the interacted items that compose the behavioral sequences for these users.}
\end{figure}
\subsection{Framework of EA-GPS}\label{subsec:framework}
\noindent In this paper, we propose a simplified graph-based solution for \ac{SR} tasks, with a dual aim of enhancing performance while remaining lightweight in its design. As shown in Fig. \ref{fig:framework}, the EA-GPS contains three key components, i.e., the graph construction, the external attentive graph convolutional network and the positional prompt-based decoder.

\subsubsection{Graph Construction}
We first construct the sequential graphs by considering two types of associations, i.e., user-item interactions and item-item sequential relationships. The resulting sequential graphs can be denoted as $\mathcal{G}=\{G_1, G_2, \dots, G_r\} $, where $r$ represents the number of training batches. 
In detail, the input Laplace matrix $\mathcal{M}\in \mathbb{R}^{(m+n)\times(m+n)}$ contains both the two-way user-item interactive relationships and the one-way item-item sequential dependency. The definitions of $\mathcal{M}$ can be formulated as:
\begin{align}
\mathcal{M}={
\left[ \begin{array}{cc}
\mathcal{A}_{I\rightarrow I} & \mathcal{A}_{U\rightarrow I}\\
\mathcal{A}_{I\rightarrow U} & \bm{0}\\
\end{array} 
\right ]}, 
\end{align}
where $\mathcal{A}_{I\rightarrow I}\in\mathbb{R}^{m\times m}$ is the weight matrix carrying the sequential dependencies from items to their neighbors; $\mathcal{A}_{U\rightarrow I}\in\mathbb{R}^{m\times n}$ is the weight matrix carrying the interactive relationships from users to items; $\mathcal{A}_{I\rightarrow U}\in\mathbb{R}^{n\times m}$ denotes the matrix which records the weights from items to users. 

\subsubsection{External Attentive Graph Convolutional Network} 
We propose an external attentive graph convolutional network, including a primary graph encoder and an external encoder. In the primary graph encoder, we merely retain the most necessary components (i.e., the neighborhood aggregating and the layer-wise sum component \cite{lightgcn20sigir}) to propagate messages on sequential graphs. In the external encoder, we utilize the External Attention (EA) mechanism to capture the global associations among nodes \cite{ea22tpami}. Compared to Self-attention (SA), EA has linear computational complexity. Moreover, the calculation of EA occurs solely between the nodes and two low-rank external memory units. Hence, we operate the external encoder in parallel, alleviating the burden on the primary graph encoder. Besides, it also allows for sharing global attention weights across different convolution layers.

\subsubsection{Positional Prompt-based Decoder} 
We present a positional prompt-based decoder to consider the complicated positional dependencies of interactions during the sequence-level representation learning. Inspired by the prompt-based methods \cite{Luoprompt2023}, an intuitive idea is to directly tell the decoder about the item positions within each sequence. Hence, we treat the absolute item positions as prompts and project them into embedding-form. Then, we construct the positional prompting templates by vertically concatenating the position embeddings with sequence embeddings. Moreover, we devise a length-adaptive sequential masking strategy to mask the prompting templates. Finally, we utilize a soft attention network to refine the sequence representations. By forcing the model to predict the masked elements based on the context, such a component is able to capture the complicated positional dependencies of items and the sequential behavioral patterns of users. \newline 

Finally, by concatenating the embeddings of users and sequences, EA-GPS results in user-specific sequential representations for final predictions.

\subsection{External Attentive Graph Convolutional Network}
\subsubsection{Primary Graph Encoder.}
In the primary graph encoder, we merely retain the neighborhood aggregation component to realize lightweight node representation learning. It learns the node representations by propagating messages on sequential graphs. Both the sequential relationships and the collaborative filtering signals are taken into account via the layer-wise aggregating protocol \cite{lightgcn20sigir}. We conduct a $\eta$ layers graph convolution on sequential graphs, where $\eta$ is a hyper-parameter that controls the layer numbers. The node representation learning on $l$-th ($0<l<\eta$) layer can be formulated as:
\begin{align}
& \hat{\bm{E}}^{(l)}=H(\hat{\bm{E}}^{(l-1)}, \mathcal{G}),
\end{align}
where $\hat{\bm{E}}^{(l)}\in \mathbb{R}^{(m+n)\times d}$ denotes the node representations on $l$-th layer, $\hat{\bm{E}}^{(l-1)}$ is that of the previous layer, $H(\cdot)$ represents the message aggregating function that calculates the average value according to neighboring nodes. $H(\cdot)$ can be further denoted as:
\begin{align}
H(\hat{\bm{E}}^{(l-1)}, \mathcal{G})= (\bm{D^{-\frac{1}{2}}} (\mathcal{M}+\bm{I}) \bm{D^{-\frac{1}{2}}}) \hat{\bm{E}}^{(l-1)},
\end{align}
where $\bm{I}\in \mathbb{R}^{(m+n)\times(m+n)}$ denotes the characteristics of nodes themselves, and $\bm{D}\in \mathbb{R}^{(m+n)\times(m+n)}$ is the normalized matrix. As the number of neighbors of each node is inconsistent, we introduce the normalized matrix $\bm{D}$ to alleviate this impact.

Then, the representations of users $\hat{\bm{E}}_{U}^{(l)}$ and items $\hat{\bm{E}}_{S}^{(l)}$ on $l$-th layers can be learned by following functions:
\begin{align}
    \hat{\bm{E}}_{U}^{(l)}=\sum_{u_k\in \mathcal{U}}\sum_{v_i\in S_k}\frac{1}{\sqrt{\lvert \mathcal{U}\rvert} \sqrt{\lvert S_k\rvert}}\hat{\bm{E}}^{(l-1)}_{v_i\rightarrow u_k}; \qquad \qquad \quad\\
    \hat{\bm{E}}_{S}^{(l)}=\sum^{\mathcal{S}}_{S_k}\left(\sum_{v_i\in S_k}\sum_{u_k\in \mathcal{U}}\frac{1}{\sqrt{\lvert S_k\rvert} \sqrt{\lvert \mathcal{U}\rvert}}\hat{\bm{E}}^{(l-1)}_{u_k\rightarrow v_i} + \sum_{v_i\in S_k}\frac{1}{\lvert S_k\rvert}\hat{\bm{E}}^{(l-1)}_{{v_{(i-1)}}\rightarrow v_i}\right),
\end{align}
where $\hat{\bm{E}}_{v_i\rightarrow u_k}$ denotes the passing message from items to users, $\hat{\bm{E}}_{u_k\rightarrow v_i}$ is the messages transferred from users to items, and the $\hat{\bm{E}}_{{v_{(i-1)}}\rightarrow v_i}$ represents the sequential information passing between neighbor items. 

Inspired by He et al. \cite{lightgcn20sigir}, we just keep the most necessary components in the primary graph encoder to maintain the operation of graph convolution. However, relying solely on the primary encoder cannot obtain global associations between nodes, which makes it difficult to extract sequential dependencies between items. 
Hence, we develop a lightweight external encoder, which runs in parallel with the primary graph encoder. 

\subsubsection{External Encoder. \label{subsubsec:external}}
\noindent In this section, we first revisit self-attention and linear attention mechanism before introducing the external encoder (We take $S_k$ as an example, $S_k$ is one of the sequences that composes the input graphs $\mathcal{G}$). 

\begin{figure}[b]
\centering
\includegraphics[width=1.0\columnwidth]{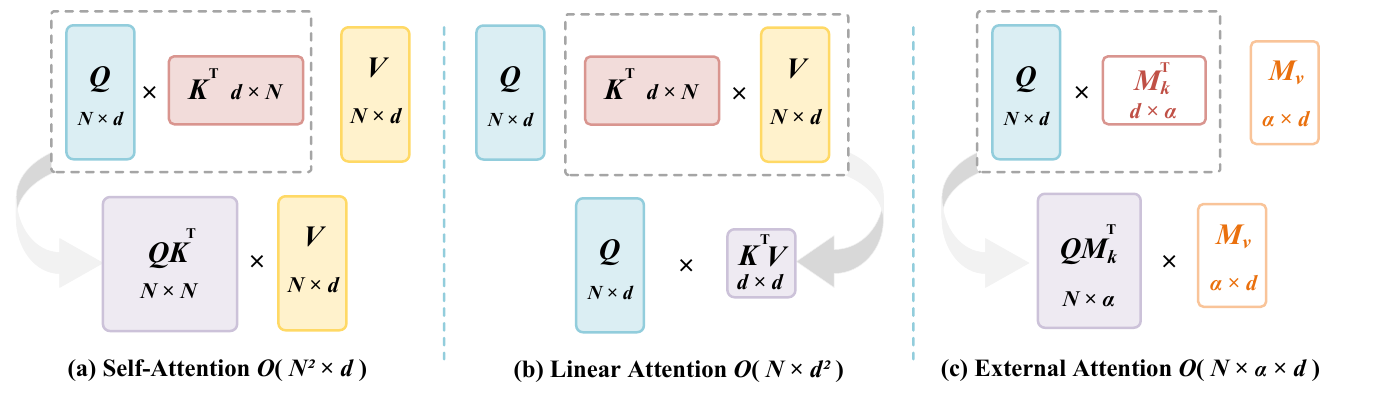}
\caption{The differences between Self-Attention (SA), Linear Attention (LA) and External Attention (EA).}
\label{fig:attention}
\Description{The differences between Self-Attention (SA), Linear Attention (LA) and External Attention (EA).}    
\end{figure}

\paragraph{\textbf{Self-Attention.}}
As a common practice in global weighting, Graph Transformer \cite{LiXRY0023} has achieved state-of-the-art performance on modeling nodes' sequential dependencies. The most essential component in Graph Transformer is the Self-Attention (SA) mechanism (as shown in Fig.~\ref{fig:attention} (a)). 
Given the input feature map $\bm{E}_{S_k} \in \mathbb{R}^{t_k\times d}$ of the sequence $S_k$, where $t_k$ denotes the number of interacted items in $S_k$ (i.e., the length of $S_k$) and $d$ is the embedding size. Self-attention mechanism maps the input features to a query matrix $\bm{Q} \in \mathbb{R}^{t_k\times d}$, a key matrix $\bm{K} \in \mathbb{R}^{t_k\times d}$, and a value matrix $\bm{V} \in \mathbb{R}^{t_k\times d}$. The calculation can be formulated as:

\begin{align}
\bm{A}_{S} = a^{S_k}_{i,j} = \text{softmax}\left(\frac{\bm{Q} \bm{K}^T}{\sqrt{d}}\right); \label{eq:self_7}\\
{\bm{Z}}_{S_k} = \bm{A}_S \bm{V}, \qquad
\end{align}
where $\bm{A}_{S} \in \mathbb{R}^{t_k \times t_k}$ is the attention map, $a^{S_k}_{i,j}$ denotes the pairwise affinity between $i$-th and $j$-th items in sequence $S_k$, and ${\bm{Z}}_{S_k} \in \mathbb{R}^{t_k \times d}$ is the refined feature representations. 

A common strategy to simplify the self-attention mechanism involves computing the attention map directly from the input features:

\begin{align}
\bm{A}_{S} = a^{S_k}_{i,j} = \text{softmax}\left( \frac{\bm{E}_{S_k} \bm{E}_{S_k}^T}{\sqrt{d}}\right); \label{eq:self_9}\\
{\bm{Z}}_{S_k} = \bm{A}_S \bm{E}_{S_k}. \qquad
\end{align}

In this simplified variant, the attention map is obtained by computing the self similarity in the feature space, and the output is the weighted feature representations ${\bm{Z}}_{S_k}$ of the input sequences $S_k$.

The self-attention mechanism fully calculates the global relevance score of each item in the sequential graph, thereby modeling the long-term preference evolution of users \cite{Vas2017attention}. However, its high computational complexity of O($N^2 \times d$) puts a huge burden on the model training \cite{lea23dasfaa}. 

\paragraph{\textbf{Linear Attention.}}
To address this limitation, Katharopoulos et al.~\cite{KatharopoulosV020} propose Linear Attention (LA) mechanism. As shown in Fig.~\ref{fig:attention} (b), LA first calculates the outer product of the key matrix and value matrix (i.e., the $\bm{K^TV}$), and then calculates the correlation by the query matrix. As LA alters the calculation order, it successfully reduces the computational complexity to the scale of O($N \times d^2$). The calculation of LA can be formulated as:

\begin{align}
\bm{A}_{L} = a^{S_k}_{i,j} = \text{softmax}\left( \frac{\bm{E}_{S_k}^T \bm{E}_{S_k}}{\sqrt{d}}\right); \label{eq:linear_11}\\
{\bm{Z}}_{S_k} = \bm{E}_{S_k} \bm{A}_{L}, \qquad
\end{align}
where $\bm{A}_{L} \in \mathbb{R}^{d \times d}$ denotes the attention map of linear attention mechanism.

However, such changes also lead to the loss of the fine-grained information from each query, resulting in a lossy attention map $\bm{A}_{L}$ for LA. 
Besides, both SA and LA require generating key or value vectors for each input feature while calculating global correlations, leading to a hign memory consumption \cite{lea23dasfaa}.

\paragraph{\textbf{External Attention in External Encoder}}
After revisiting SA and LA, we continue to introduce the key component of EA-GPS (i.e. the external encoder), which utilizes External Attention (EA) mechanism (as shown in Fig.~\ref{fig:attention} (c)) to measure the global correlations between item nodes.
This external encoder consumes sequence embeddings from each convolutional layer as its inputs and produces output embeddings with weighted scores. We also take $S_k \in \mathcal{S}$ as an example to detail the calculation of external encoder, and the input features learned by primary graph encoder can be denoted as $\bm{E}_{S_k}$.

In contrast to self-attention and linear attention mechanism, the External Attention (EA) focuses on computing global weights between the input features $\bm{E}_{S_k}$ and two external memory units, i.e., $\bm{M_k} \in \mathbb{R}^{\alpha \times d}$ and $\bm{M_v}\in \mathbb{R}^{\alpha \times d}$ ($\alpha$ is a hyper-parameter that controls the dimension of the external memory units). The computation can be elaborated as Eqn. (\ref{eq:external_13} - \ref{eq:external_14}):
\begin{align}
\bm{A}_E = \Tilde{a}^{S_k}_{i,j} = \text{Norm}\left(\frac{\bm{E}_{S_k} \bm{M_k}^T}{\sqrt{d}}\right);\label{eq:external_13} \\
{\bm{Z}}_{S_k} = \bm{A}_E \bm{M_v},\label{eq:external_14} \qquad
\end{align} 
where $\bm{A}_E \in \mathbb{R}^{t_k \times \alpha}$ is the attention matrix, $\Tilde{a}^{S_k}_{i,j}$ in Eqn. (\ref{eq:external_13}) is the similarity between the $i$-th items in $S_k$ and the $j$-th row of $\bm{M_k}$. Here, $\bm{M_k}$ and $\bm{M_v}$ are two learnable matrices that respectively play the role as the \textbf{Q}uery and \textbf{V}alue matrices in external attention. 

As we are able to adjust the dimension of the external memory units by varying the hyper-parameter $\alpha$, the computational complexity of EA could be approximated as linear (i.e., O$(N \times \alpha \times d)$). When $\alpha$ is set to $d$, the computational complexity of EA aligns with LA (as O$(N \times d^2)$). 
Rather than generating key and value vectors for each input feature, EA only maintains two external memory units during the global weighting stage, significantly reducing the computational overhead.
Besides, since the two memory units are independent of the input features, the weights preserved within $\bm{M_k}$ and $\bm{M_v}$ could be shared across GCNs' layers, achieving better global weighting performance than SA or LA for GSRs.

In Eqn. (\ref{eq:external_13}), we replace the softmax($\cdot$) with a simple l2-normalization. The reason for adopting softmax in Eqn. (\ref{eq:self_9}) and Eqn. (\ref{eq:linear_11}) is mainly because the attention maps in SA and LA are sensitive to the scale of the input features. However, this constraint is no longer effective with the participation of external memory units, so we replace it.

Inspired by the architecture of Transformer \cite{Vas2017attention}, we convert the input feature map into $\beta$ multi-head query vectors as $\{\bm{E}_{S_k}^{(1)}, \bm{E}_{S_k}^{(2)}, \dots, \bm{E}_{S_k}^{(\beta)}\}$, where $\beta$ is a hyper-parameter that controls the number of heads. 
Similarly, we define the multi-head weighted representations as $\{ \bm{Z}_{S_k}^{(1)}, \bm{Z}_{S_k}^{(2)}, \dots, \bm{Z}_{S_k}^{(\beta)} \}$. Hence, the final refined embedding $\bm{\hat{Z}_{S_k}}$ for $S_K$ can be further denoted as:
\begin{align}
\bm{\hat{Z}}_{S_k}=Concat\left(\bm{Z}_{S_k}^{(1)}, \bm{Z}_{S_k}^{(2)}, \dots, \bm{Z}_{S_k}^{(\beta)}\right) \cdot \bm{W}_1,
\end{align}
where $\bm{W}_1$ represents a linear transformation matrix that aligns the dimensions of inputs and multi-heads outputs. Thus, the weighted representations of items across all sequences are expressed as:
\begin{align}
    \bm{\hat{Z}}_S=\sum_{S_k}^{\lvert \mathcal{S}\rvert} \bm{\hat{Z}}_{S_k}.
\end{align}

\subsubsection{Message aggregation between encoders}

To aggregate the messages from encoders on $l$-th layer, EA-GPS appends the global correlations from EA to the item representations $\bm{E}_{S}^{(l)}$ via residual links, the updated representations are formulated as:  
\begin{align}
    \quad\bm{E}_{S}^{(l)}=\text{LayerNorm}(\hat{\bm{E}}_{S}^{(l)} + \delta \bm{\hat{Z}}_S^{(l)}),  
\end{align}
where $\delta$ is the hyper-parameter controls the participation of the correlations generated by the external encoder. After extensive experiments, we found that the optimal performance is always achieved when the $\delta=1$. Hence, we did not report the impact of the $\delta$ in this paper, but set it directly to 1.

In the primary graph encoder, the node embeddings at 0-th layer ($\bm{E}_{U}^{(0)}$ and $\bm{E}_{S}^{(0)}$) are optimized individually, with which the node representations at higher layers are calculated by layer-wise message passing strategies. Then, the final node representations are learned via layer combination, which are formulated as:
\begin{align}
    \bm{E}_{U}=\sum_{l=0}^{\eta} \frac{1}{1+\eta}\bm{E}_{U}^{(l)};\quad \bm{E}_{S}=\sum_{l=0}^{\eta} \frac{1}{1+\eta}\bm{E}_{S}^{(l)},
\end{align}
where $\eta$ is a hyper-parameter that controls the number of graph convolution layers, $\bm{E}_{U}$ and $\bm{E}_{S}$ denote the resulting node representation for users and items, respectively.

In this way, the EA-GPS facilitates the detection of the global correlations among items with a lower computational complexity. Then we present a positional prompt-based decoder for sequence-level representation learning.

\subsection{Positional Prompt-based Decoder} \label{subsec:prompt}
\noindent At the stage of sequence representation learning, the traditional relative position encoding in the Graph Transformer \cite{XiaHXP23} fails to consider the complicated positional dependencies of items (e.g., items' long-term or discontinuous relationships) \cite{WuPCFC21}. The primary reason is that the relative positional relationships between adjacent items are unable to assist models for perceiving the position of an item within the entire sequence.
Inspired by the prompt-based methods \cite{Luoprompt2023}, we have an intuitive idea that directly tells the model about the item positions. Specifically, we treat the absolute item position as prompts and then propose a positional prompt-based decoder to assist the sequence representation learning. The input of this decoder is the node representations (i.e., $\bm{E}_{S}$ and $\bm{E}_{U}$) and the item positions of each sequence ($\mathcal{T}$). The positional prompt-based decoder contains three key parts, i.e., prompting templates, sequential masking, and soft attention network.

\begin{figure}[t]
    \centering
    \includegraphics[width=0.95\columnwidth]{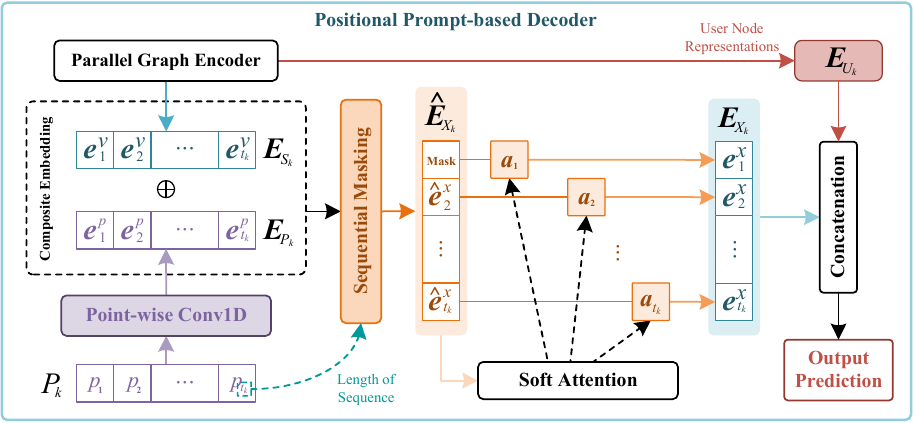}
    \caption{The architecture of the positional prompt-based decoder (taking $S_k$ as an example).}
    \label{fig:prompts}
    \Description{The architecture of the positional prompt-based decoder (taking $S_k$ as an example).}  
\end{figure}

\subsubsection{The Construction of Prompting Template} 
We also take $S_k$ as an example, its sequence embedding is $\textbf{E}_{S_k} \in \mathbb{R}^{t_k \times d}$. As shown in Fig. \ref{fig:prompts}, we treat the the exact position of each item in sequences $P_k=\{p_1, p_2, \dots, p_{t_k}\}$ as the positional prompt. We initilaize it to embedding-form as $\hat{\bm{E}}_{P_k}\in \mathbb{R}^{{t_k} \times d_1}$. Then, we linearly project $\hat{\bm{E}}_{P_k}$ to the same dimension as $\bm{E}_{S_k}$ via point-wise Conv1D. For linear dimensional transformation, both the kernel size and stride of the Conv1D are set to 1 \cite{Wu_Capsule_Graph_23_TNNLs}. The calculation is denoted as Eqn. (\ref{eqn:conv1d}).
\begin{equation}
    \mathbf{E}_{P_k} = \hat{\bm{E}}_{P_k} \ast \mathbf{W}_c + \mathbf{b}_c, \label{eqn:conv1d}
\end{equation}
where $\ast$ denotes the 1D convolutional operation, $\mathbf{W}_c \in \mathbb{R}^{d_1 \times d}$ represents the convolutional kernel, $\mathbf{b}_c \in \mathbb{R}^{1 \times d}$ is the bias term. 

After obtaining the embedding-form positional prompts, we construct the prompting template by vertically concatenating the positional prompts $\bm{E}_{P_k}$ with the sequence embeddings $\bm{E}_{S_k}$ (As shown in Fig. \ref{fig:prompts}). The main purpose of vertical concatenation is to integrate positional information into the sequence representation, enabling the model to better capture long-term dependencies and contextual relationships within the sequence. By incorporating absolute positional prompts, the model is able to understand the temporal evolution of user preferences more effectively, which is crucial for sequential recommendation tasks. Besides, compared to the dimension of the conventional horizontal concatenation $\mathbb{R}^{(t_k + t_k)\times d}$, the prompting template by vertical concatenation has a dimension of $\mathbb{R}^{t_k \times 2d}$. Since the embedding-size $d$ is usually small (i.e., below 128), there is no significant difference between them in computational complexity.

Subsequently, we employ a Feedforward Neural Network (FNN) to map the concatenated template into sequence-level representation space:
\begin{align}
    \bar{\bm{E}}_{X_k}=\text{ReLU}(\bm{W}_2\left[\begin{array}{c}
         \bm{E}_{S_k}  \\
         \oplus \\
         \bm{E}_{P_k} 
    \end{array} \right]+\bm{b}_2),
\end{align}
where the $\bar{\bm{E}}_{X_k}=\left\{\bar{\bm{e}}^x_1, \bar{\bm{e}}^x_2, \dots, \bar{\bm{e}}^x_i, \dots, \bar{\bm{e}}^x_{t_k}\right\}$ is the representation of the positional prompting template, $\bm{W}_2$ and $\bm{b}_2$ are trainable matrix and bias vector. 

Notably, when handling exceptionally lengthy sequences, the dimension of the prompting template may become extremely large. A common practice is to segment such extended sequences into several sub-sequences for the same user during the data pre-processing \cite{HeHSLJC18, ZhangLXXLHCM22, LiCZY21}. However, this strategy struggles to preserve the entire sequential dependencies from the original sequence, potentially leading to the loss of long-range correlations. To address this limitation, the external encoder we have introduced in Section \ref{subsubsec:external} is capable of considering the global dependency relationships across different sequences. Consequently, EA-GPS could handle the exceptionally lengthy sequences confidently without the risk of compromising performance.

Besides, we validate the effectiveness of this prompting template construction strategy by conducting additional ablation experiments in Section \ref{subsec:ablation}. We compare EA-GPS with a variant method named GPS$_{RPE}$ that adopts relative position encoding for sequence representation learning. Experimental results demonstrate the superiority of our solutions on SR tasks.

\subsubsection{Sequential Masking}
Based on the initialized representation of prompting templates, EA-GPS employs a sequential masking strategy to adaptively foster the model to better understand the sequential context. 
Specifically, rather than randomly determining whether to mask each element in the prompting template, we randomly mask $t_k \times \gamma$ elements on the positional prompting template, where $\gamma$ is a hyper-parameter that controls the proportion of items to be masked in the prompting templates. We define the masking vector as $[\bm{m}_1, \bm{m}_2, \dots, \bm{m}_i, \dots, \bm{m}_{|t_k|}]$, where $\bm{m}_i$ denotes a binary value indicating whether the corresponding element in the template should be masked. Then, the masking operation is denoted as:
\begin{align}
    \bm{m}_i =   \begin{cases} 
            1 & \text{if } \bar{\bm{e}}^x_i \text{ is to be masked,} \text{ and } g_k \leq t_k \times \gamma; \\
            0 & \text{otherwise},
            \end{cases}
\end{align}
where $\bar{\bm{e}}^x_i$ is an element of the prompting templates, $g_k$ denotes the number of elements that has already been masked in the prompting templates $\bar{\bm{E}}_{X_k}$.
Intuitively, the masked embedding $\hat{\bm{E}}_{X_k}=\left\{\hat{\bm{e}}^x_1, \hat{\bm{e}}^x_2, \dots, \hat{\bm{e}}^x_i, \dots, \hat{\bm{e}}^x_{t_k}\right\}$ of prompting template is denoted as:
\begin{align}
    \hat{\bm{e}}^x_i = \bm{m}_i \cdot \bar{\bm{e}}^x_i + (1 - \bm{m}_i) \cdot \text{[mask]}. \label{eq:sequential_mask}
\end{align} 

In conventional masking strategies, the quantity of masked elements within each prompting template is unequal. Short sequences may inadvertently mask critical information when the masking is too extensive, compromising their representations. Conversely, long sequences may not leverage the full potential of prompts if the masking is too limited. By carefully considering sequence length during masking, sequential masking in EA-GPS ensures equitable treatment of sequences with varying lengths.

\subsubsection{Soft Attention Network}
After obtaining the masked templates $\hat{\bm{E}}_{X_k}$, we devise a soft attention network to activate the masked prompting template. Here, we leverage the last element $\hat{\bm{e}}^x_t$ within template as additional supervision signal, guiding the decoder to refine the sequence representation learning. For the $i$-th element $\hat{\bm{e}}^x_i$ in $\hat{\bm{E}}_{X_k}$, we compute its relevance to $\hat{\bm{e}}^x_t$ by Eqn. (\ref{eqn:att_1}).
\begin{equation}
    a_{i,t} = \frac{\exp{\left(f(\hat{\bm{e}}^x_i, \hat{\bm{e}}^x_t)\right)}}{\sqrt{\sum^{|S_k|}_{i=1}\exp{\left(f(\hat{\bm{e}}^x_i, \hat{\bm{e}}^x_t)\right)} }},\label{eqn:att_1}
\end{equation}
where $f(\cdot)$ is the cosine similarity function that calculates the correlations between elements, $a_{i,t}$ denotes the resulted correlations.

Then, the refined sequence representation $\bm{E}_{X_k}$ is obtained by Eqn. (\ref{eqn:att_2}) as: 
\begin{equation}
    \bm{E}_{X_k} = \sum^{|S_k|}_{i=1} a_{i,t} \hat{\bm{e}}^x_i. \label{eqn:att_2}
\end{equation}

By employing such soft attention mechanism, the model is forced to infer the masked element based on the contextual information and the positional prompts. It facilitates the model to better understand the long-term positional dependencies and contextual relationships within sequences.

Finally, the user-specific sequence-level representations $\bm{H}_{S_k}$ for user $u_k$ are learned by concatenating refined sequence embeddings $\bm{E}_{X_k}$ with the user embeddings $\bm{E}_{u_k}$ as:
\begin{align}
    \bm{H}_{S_k} = \text{Concat}[\bm{E}_{X_k},\bm{E}_{u_k}].
\end{align}

The representation learning process for the single user $u_k$ can also be easily extended to all users. Hence, the final representation output by the positional prompt-based decoder can be denoted as:
\begin{align}
    \bm{H}_{S} = \text{Concat}[\bm{H}_{S_1}, \bm{H}_{S_2}, ..., \bm{H}_{S_k}, ..., \bm{H}_{S_n}].
\end{align}


\subsection{Final Prediction}\label{subsec:prediction}
\noindent After the sequence representation learning, EA-GPS generates the user-specific sequence-level representations $\bm{H}_{S}$. We feed it into the prediction layer to get the final probabilities.
\begin{align}
P(I_{i+1}|S, U) = softmax(\bm{W}_6\cdot\bm{H}_{S}^\mathrm{T}+\bm{b}_6),
\end{align}
where $\bm{W}_6$ is a learnable weighting matrix, $\bm{b}_6$ is a bias term. 

Finally, the cross-entropy loss function is adopted to optimize the parameters in EA-GPS, as shown in Eqn. (\ref{equ:loss}):
\begin{align}
\mathcal{L}_S = -\frac{1}{|\mathcal{S}|}\sum_{S \in \mathcal{S}}\sum_{I_i \in S}\text{log} P(I_{i+1}|S, U).
\label{equ:loss}
\end{align}

\section{Experiments}
\noindent We conduct extensive experiments on five real-world datasets to verify the effectiveness of our proposed method. We first introduce the experimental settings, and then analyze the performance of EA-GPS to answer the following \textbf{R}esearch \textbf{Q}uestions.

\begin{itemize}
    \item[\textbf{RQ1}] How does EA-GPS's lightweight design perform in terms of training efficiency and parameter scale?
    \item[\textbf{RQ2}] How does our proposed EA-GPS perform, compared to other state-of-the-art sequential recommenders?
    \item[\textbf{RQ3}] How do the key components of EA-GPS (i.e., external encoder and positional prompt-based decoder) contribute to the recommendation performance? 
    \item[\textbf{RQ4}] How do the hyper-parameters affect the performance of EA-GPS?
\end{itemize}

\subsection{Experimental Settings}

\begin{table}[t]
    \centering
    \normalsize
     \caption{Statistics of five real-world datasets.}
    \begin{tabular}{lccccc}
    \toprule
    \midrule
    \multicolumn{1}{c}{\textbf{Dataset}}&\multicolumn{1}{c}{\textbf{FOOD}}&\multicolumn{1}{c}{\textbf{MOVIE}}&\multicolumn{1}{c}{\textbf{BOOK}}&\multicolumn{1}{c}{\textbf{MOVIELENS}}&\multicolumn{1}{c}{\textbf{DOUBAN}} \\
    \midrule
    Items &14,636 &67,141 &126,547 &8,674 &61,362 \\
    Interactions &607,523 &978,226 &1,678,006 &1,507,131 &1,024,445  \\
    Users & 6,582 &11,459 &13,724 &55,545 &9,211 \\
    \midrule
    Training Sequences &29,444 &81,455 &76,146 &62,653 &84,029  \\
    Testing Sequences &12,618 &34,909 &32,633 &15,665 &21,008  \\
    \midrule
    \bottomrule
    \end{tabular}
    \label{tab:dataset_statistics}
\end{table}
\subsubsection{Datasets} \label{datasets} 
As illustrated in Table \ref{tab:dataset_statistics}, five real-world datasets (i.e., FOOD, MOVIE, BOOK, MOVIELENS, and DOUBAN) are used in experiments to evaluate the performance of EA-GPS. 

FOOD, MOVIE, and BOOK are gathered from the Amazon online platform\footnote{http://jmcauley.ucsd.edu/data/amazon} by Fu et al. \cite{fu2019ccr}. These product review datasets record users’ review behaviors of three different channels (i.e., amazon-food, amazon-movie, and amazon-book) from May 1996 to July 2014. The FOOD dataset comprises Amazon users’ consumption and rating interactions on food, consisting of 6,582 users, 42,062 sequences, and 607,523 interactions. The MOVIE dataset documents users’ movie watching and rating activities, containing 11,459 users, 116,364 sequences, and 978,226 interactions. The BOOK dataset records users’ book reading and rating behaviors, containing 13,724 users, 108,779 sequences, and 1,678,006 interactions. 

MOVIELENS dataset is selected from the MovieLens-10M public dataset, which is a widely used dataset for recommender systems and collaborative filtering research \cite{quadrana2017hrnn, HeHSLJC18, SunLWPLOJ19} and is collected by GroupLens\footnote{https://grouplens.org/datasets/movielens/}. The original MovieLens-10M contains approximately ten million ratings from 72,000 users on 11,000 movies. Each user has rated at least twenty movies, and the ratings are on a scale from 1 to 5. Since the Sequential Recommendation tasks only focus on the interactions and their sequential orders, we only keep those interactions with a rating of 5 (the most positive feedback from users) and then generate sequences for each user based on the timestamps of each interaction \cite{SunLWPLOJ19}. Thus, the resulting MOVIELENS dataset contains 55,545 users, 78,318 sequences, and 1,507,131 interactions.

DOUBAN is a book review dataset collected from one of the most popular Chinese media platforms (i.e., Douban\footnote{https://book.douban.com/}) by Zhang et al. \cite{lea23dasfaa}. It contains 9,211 users, 105,037 sequences, and 1,024,445 interactions.

For each dataset, we randomly select 80$\%$ of the historical sequences as training sets and the remaining 20$\%$ as testing sets. We split each sequence into small fragments within one year to facilitate the model training. To avoid the impact of cold items, we drop those short sequences with fewer than three interactions and filter out items that have interacted fewer than five times. 
\subsubsection{Baseline Methods} \label{baselines}
To verify the performance of EA-GPS, we select thirteen state-of-the-art methods belonging to three different categories as baselines. The details of each method are shown in Table \ref{tab:baselines}. Note that, we turn the original graph collaborative filtering methods (i.e., NGCF, LightGCN, and SGL) into graph-based sequential recommendation baselines by feeding them with sequential graphs. Hence, these methods are able to simultaneously consider the user-item interactions and sequential relationships between items.
\begin{table}[t]
\centering
\caption{The description of the compared baseline methods.}
\label{tab:baselines}
\footnotesize
\begin{tabular}{p{3cm}|p{2.5cm}|p{7cm}}
\toprule
\midrule
\textbf{Category} & \textbf{Method} & \textbf{Descriptions} \\ 
\midrule
\textbf{Traditional \qquad \qquad Recommendation} & \textbf{BPR-MF (2009)} & This is a traditional recommendation model, exploits matrix factorization to predict user-item interactions \cite{koren2009matrix}. \\
\cmidrule{2-3}
& \textbf{NCF (2017)} & This traditional method uses deep neural networks to model the collaborative filtering signals from items \cite{he2017ncf}. \\
\midrule
\textbf{Sequential \qquad \qquad \quad Recommendation} & \textbf{GRU4REC (2016)} & This is an early proposed RNN-based SR method, which exploits GRU to model the sequential dependencies \cite{hidasi2016srnn}. \\
\cmidrule{2-3}
& \textbf{NAIS (2018)} & This is an attention-based SR solution, which incorporates a nonlinear attention network to measure the similarities between items \cite{HeHSLJC18}. \\
\cmidrule{2-3}
& \textbf{SASRec (2018)} & A classic attention-based sequential recommendation algorithm, which employs self-attention mechanisms to model user-item interactions \cite{KangM18}. \\
\cmidrule{2-3}
& \textbf{Bert4Rec (2019)} & This method leverages BERT to capture complicated dependencies among items from user behavior sequences \cite{SunLWPLOJ19}. \\
\midrule
\textbf{Graph-based \qquad \qquad \quad Sequential \qquad \qquad \quad Recommendation} & \textbf{NGCF (2019)} & This is a graph collaborative filtering method that employs multi-layers Graph Convolutional Networks (GCNs) to capture higher-order connectivity within graphs \cite{ngcf19sigir}. \\
\cmidrule{2-3}
& \textbf{LightGCN (2020)} & LightGCN further simplifies the NGCF by removing the feature transformation matrix and nonlinear activation functions \cite{lightgcn20sigir}. \\
\cmidrule{2-3}
& \textbf{SGL (2021)} & This is a self-supervised graph collaborative recommender based on contrastive graph augmentations. \cite{WuWF0CLX21}. \\
\cmidrule{2-3}
& \textbf{SURGE (2021)} & SURGE is a GSR method that constructs item-item interest graphs and learns sequential representations via cluster-aware and query-aware graph convolutions \cite{ChangGZHNSJ021}.\\
\cmidrule{2-3}
& \textbf{GCL4SR (2022)} & This GSR method leverages a weighted item transition graph to consider global context information and relieve the noisy impact in the sequence data \cite{ZhangLXXLHCM22}.\\
\cmidrule{2-3}
& \textbf{PTGCN (2023)} & PTGCN is a positional information-enhanced, time-aware graph convolutional network, learning dynamic representations via a self-attention mechanism on bipartite graphs \cite{HuangMLDWL23}. \\
\cmidrule{2-3}
& \textbf{TGT (2023)}  & This is a Transformer-based GSR method, which leverages the positional encoding to consider the relative position of items within sequences \cite{XiaHXP23}.\\ 
\midrule
\bottomrule
\end{tabular}
\end{table}

\subsubsection{Evaluation Metrics} \label{evaluation} 
In every dataset, we hold out the last observed items in each sequence as the target items (a.k.a, ground truth items). To evaluate the performance of EA-GPS and other baseline methods, we report the top-N results on two commonly used metrics, i.e., Recall@N and MRR@N \cite{hidasi2016srnn,tidagcn2022} (we set $N=\{5, 10\}$ in experiments).
\begin{itemize}
    \item {\textbf{Recall@N}}. Recall is a popular evaluation metric for recommendation systems, measuring the proportion of relevant items successfully retrieved within the top-N recommendations. Recall@N is well-suited for scenarios where the user is interested in discovering new items or exploring a wide range of options, such as e-commerce, video streaming, or music recommendation platforms. However, it may not be suitable for situations where the user is primarily interested in a small number of highly relevant items, as Recall@N does not account for the ordering of recommendations \cite{LiCZY21}.
    \item {\textbf{MRR@N}}. Mean Reciprocal Rank (MRR) is another widely used evaluation metric for recommendation systems. It calculates by taking the average of the reciprocal ranks of the relevant items in a list of recommendations, truncated at a specific depth N. It is particularly suitable for scenarios where the goal is to identify a small set of relevant items from a large pool of candidates. Compared with Recall@N, MRR@N incorporates the rank of items, making it a crucial metric in recommendation tasks where the order of item rankings is significant \cite{tidagcn2022}.
\end{itemize}

\subsubsection{Implementation Details} \label{implementations}
We implement EA-GPS with TensorFlow and accelerate the training process via NVIDIA RTX 3090 (24G) GPU. We exploit Xavier \cite{glorot2010xavier} for parameter initialization and employ Adam \cite{kingma2014adam} to optimize them.
For training the model, we set the batch-size as 256, the dropout ratio as 0.1, and the learning rate as 0.0001 for FOOD and MOVIELENS, 0.001 for BOOK and DOUBAN, 0.003 for MOVIE, respectively. For EA-GPS, we varied the embedding size from 16 to 128 to evaluate its lightweight performance. We set the regularization coefficient as 1e-7 for the external encoder. The hyper-parameter $\alpha$ controls the dimension of the external memory units. We search it within the range $\{ 8, 16, 32, 64, 128\}$. $\beta$ controls the number of attention heads. We investigate it within the range $\{1, 2, 4, 8, 16\}$. $\gamma$ controls the proportion of items in a single sequence that are masked. We search its value in [0, 1] with a step size of 0.1. $\eta$ controls the number of convolution layers. Setting $\eta$ too high can lead to a significant over-smoothing problem for graph convolutional networks. Hence, we follow the parameter settings of TGT and set $\eta$ to 2. To ensure the fairness of experiments, we also unify the number of layers for other GCN-based baselines to 2 layers. Note that, the embedding-size is set to 16 uniformly for EA-GPS and other baseline models to ensure a fair comparison. For other hyper-parameters of baselines, we initially adopt the best settings reported in their paper and then fine-tuned them on each dataset. 

\begin{figure}[b]
    \centering
    \includegraphics[width=1.0\columnwidth]{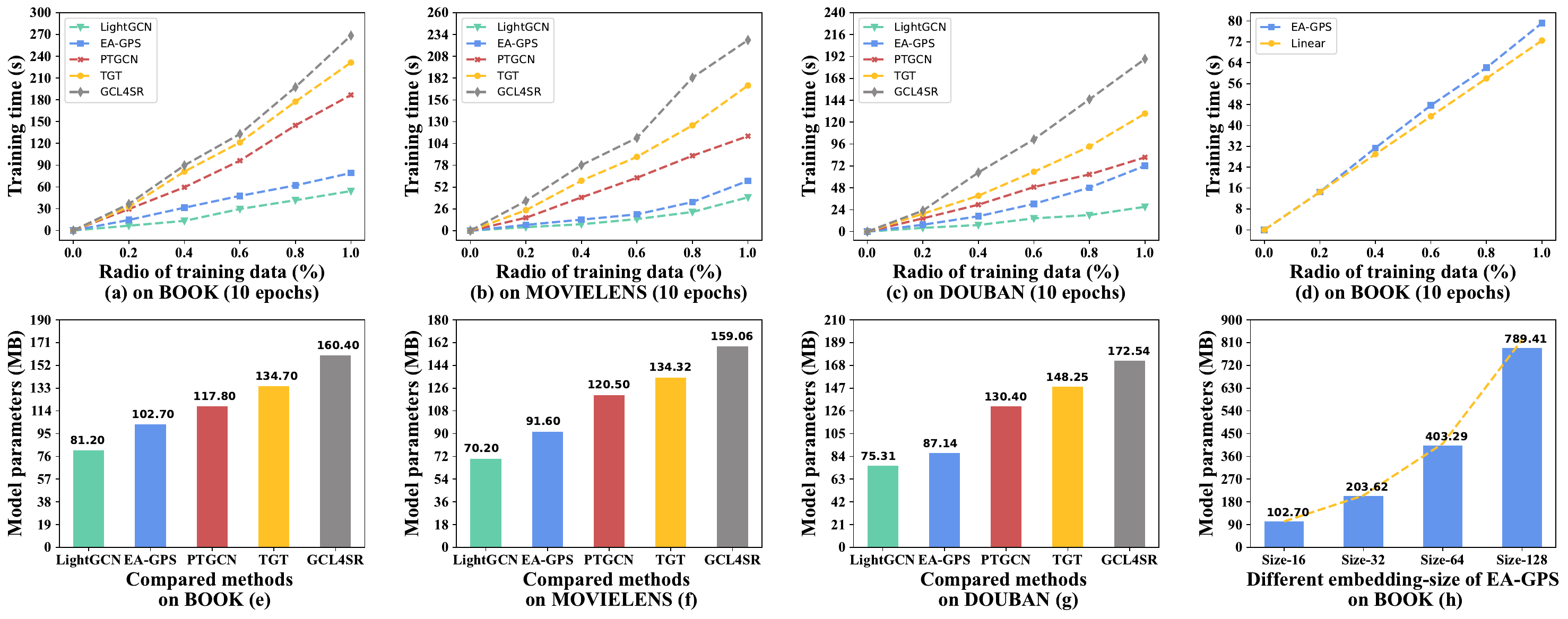}
    \caption{Time consumption and the parameter scale of EA-GPS compared with LightGCN, TGT, PTGCN and GCL4SR.}
    \label{fig:model_efficiency}
    \Description{Time consumption and the parameter scale of EA-GPS compared with LightGCN, TGT, PTGCN and GCL4SR.}  
\end{figure}

\subsection{Parameter Scale and Training Efficiency (RQ1) \label{subsec:training}}
\noindent In this section, we compare EA-GPS with the other four most competitive GSR baseline methods (i.e., LightGCN, TGT, PTGCN and GCL4SR) in terms of time consumption and parameter scale. The corresponding comparative experimental results are reported in Fig. \ref{fig:model_efficiency}.

\subsubsection{Time Consumption} 
We assess the time consumption of model training by varying the ratio of input data from 0.2 to 1.0 on three datasets (i.e., BOOK, MOVIELENS, and DOUBAN). The results are shown in Fig. \ref{fig:model_efficiency}, from which we can clearly get the following observations. 1) From Fig. \ref{fig:model_efficiency} (a) to (c), it's evident that EA-GPS requires less training time compared to other state-of-the-art GSR methods (i.e., PTGCN, TGT, and GCL4SR), and even approaches the efficiency of LightGCN (i.e., a GSR with only the collaborative filtering component). This observation demonstrates the effectiveness of our proposed external attentive graph convolutional network and positional prompt-based decoder in improving the training efficiency of GSRs. 2) EA-GPS outperforms the Transformer-based method (i.e., the TGT) on the training time consumption, demonstrating the external attention-based graph encoder in EA-GPS has better lightweight performance than that in the graph Transformer. 3) EA-GPS requires less training time consumption compared to the position information-enhanced GSRs based on the relative positional encoding, i.e., PTGCN and TGT. This observation demonstrates that our proposed positional prompt-based decoder has better training efficiency than the decoder in the graph Transformer.

To evaluate the scalability of EA-GPS, we conduct experiments on the BOOK dataset and compared the results with a linear reference line in Fig. \ref{fig:model_efficiency} (d), which represents the ideal training consumption. This result reveals that the training cost of EA-GPS is nearly linearly associated with the ideal consumption, suggesting that EA-GPS is scalable for large-scale datasets.

\subsubsection{Parameter Scale} We also conduct a comparison of parameter scales among the above GSRs. As shown in Fig. \ref{fig:model_efficiency} (e) to (g), EA-GPS has far fewer parameters than PTGCN, TGT and GCL4SR, and its parameter count is almost equal to the LightGCN. This result validates EA-GPS’s superior performance in conserving computational resources. Furthermore, it suggests that EA-GPS is more suitable for deployment on resource-constrained edge devices compared to the current state-of-the-art GSRs.

To further illustrate the lightweight performance of EA-GPS, we tested various embedding sizes on the BOOK dataset, as demonstrated in Fig. \ref{fig:model_efficiency} (h). The experimental results demonstrate that the parameter scale of our graph-based solution increases approximately linearly with different embedding sizes, providing further support for RQ1.

\begin{table*}[b]
  \centering
  \normalsize
   \caption{Experimental results ($\%$) for different methods on five real-world datasets. The best results are indicated in bold, while underlined values indicate the sub-optimal results.}
   \resizebox{\linewidth}{!}{ 
    \begin{tabular}{l|c|cc|cccc|ccccccc|c}
    \toprule
    \midrule
    \multicolumn{1}{c|}{\textbf{Dataset}} & \multicolumn{1}{c|} {\textbf{Metric}} & \multicolumn{1}{c}{BPR-MF} & \multicolumn{1}{c|}{NCF} & \multicolumn{1}{c}{GRU4REC} & \multicolumn{1}{c}{NAIS} & \multicolumn{1}{c}{SASRec} & \multicolumn{1}{c|}{BERT4Rec} & \multicolumn{1}{c}{NGCF} & \multicolumn{1}{c}{LightGCN} & \multicolumn{1}{c}{SGL} & \multicolumn{1}{c}{SURGE}& \multicolumn{1}{c}{GCL4SR}& \multicolumn{1}{c}{\textbf{PTGCN}} & \multicolumn{1}{c|}{TGT} & \multicolumn{1}{c}{\textbf{EA-GPS}}\\
    \midrule
    \multicolumn{1}{c|}{\multirow{4}[1]{*}{\textbf{FOOD}}}& \multicolumn{1}{c|}{Recall@5} &28.88 &57.23 &69.55 &71.49 &73.29 &73.65 &75.23 &74.58 &76.21 &76.16 &\underline{76.32} &76.22 &76.19 & \textbf{78.02}\\
    &  \multicolumn{1}{c|}{Recall@10} &30.16 &59.05 &70.21 &72.80 &74.96 &75.33 &76.36 &75.66 &78.45 &78.36 &\underline{78.54} &78.22&78.12 & \textbf{79.26}\\
    &  \multicolumn{1}{c|}{MRR@5} &29.79 &55.48 &67.27 &69.06 &71.23 &71.59 &73.36 &73.10 &74.33 &74.29 &\underline{74.38}& \underline{74.38} &74.35 & \textbf{75.63}\\
    &  \multicolumn{1}{c|}{MRR@10} &31.24 &55.82 &69.08 &69.97 &72.48 &72.83 &73.91 &73.29 &74.61 &74.51 &\underline{74.68} & 74.60 &74.41 & \textbf{75.80}\\
    \midrule
    \multicolumn{1}{c|}{\multirow{4}[1]{*}{\textbf{MOVIE}}}& \multicolumn{1}{c|}{Recall@5} &12.84 &22.56 &32.44 &32.32 &32.89 &33.05 &36.83 &36.12 &36.55 &36.41 &36.68 &36.62 &\underline{36.70} & \textbf{37.33}\\
    &  \multicolumn{1}{c|}{Recall@10} &16.66 &24.01 &33.62 &34.30 &34.77 &34.94 &37.23 &36.85 &36.97 &36.80 &37.75 &37.67 &\underline{37.88} & \textbf{38.62}\\
    &  \multicolumn{1}{c|}{MRR@5} &10.32 &19.05 &27.30 &27.56 &29.00 &29.15 &30.01 &29.63 &30.06 &30.22 &30.95 &\underline{31.03} &30.94 & \textbf{32.11}\\
    &  \multicolumn{1}{c|}{MRR@10} &11.21 &19.37 &27.64 &28.13 &29.51 &29.70 &31.26 &31.00 &30.94 &31.08 &31.22 &31.19 &\underline{31.23} & \textbf{32.22}\\
    \midrule
    \multicolumn{1}{c|}{\multirow{4}[1]{*}{\textbf{BOOK}}}& \multicolumn{1}{c|}{Recall@5} &11.33 &22.57 &32.41 &32.63 &34.13 &34.30 &36.22 &36.02 &35.82 &36.01 &36.52 &36.69 &\underline{36.71} & \textbf{38.11}\\
    &  \multicolumn{1}{c|}{Recall@10} &13.52 &24.35 &33.64 &33.85 &34.75 &35.00 &36.93 &36.91 &36.54 &37.00 &37.43 & 37.62 &\underline{37.82} & \textbf{38.77}\\
    &  \multicolumn{1}{c|}{MRR@5} &9.84 &19.43 &27.99 &27.92 &29.99 &30.29 &30.82 &30.87 &30.71 &31.20 &31.55 & 31.57 &\underline{31.61} & \textbf{32.59}\\
    &  \multicolumn{1}{c|}{MRR@10} &10.27 &19.82 &28.70 &27.96 &30.31 &30.64 &31.18 &31.13 &31.07 &31.46 &31.66 & 31.71 &\underline{32.00} & \textbf{32.74}\\
    \midrule
    \multicolumn{1}{c|}{\multirow{4}[1]{*}{\textbf{MOVIELENS}}}& \multicolumn{1}{c|}{Recall@5} &2.11 &2.36 &2.59 &2.52 &2.64 &2.69 &2.88 &2.88 &2.89 &3.14 &3.66 &3.77 &\underline{3.82} & \textbf{4.47}\\
    &  \multicolumn{1}{c|}{Recall@10} &3.28 &3.36 &4.70 &4.99 &4.85 &5.25 &4.88 &4.87 &5.98 &6.01 &6.32 & \underline{6.40} &6.25 & \textbf{7.87}\\
    &  \multicolumn{1}{c|}{MRR@5} &1.03 &1.32 &1.46 &1.55 &1.53 &1.65 &1.72 &1.76 &1.88 &1.96 &1.99 & 2.00&\underline{2.02} & \textbf{2.19}\\
    &  \multicolumn{1}{c|}{MRR@10} &1.07 &1.35 &1.49 &1.58 &1.65 &1.82 &1.92 &1.99 &2.04 &2.00 &2.11 &\underline{2.14} &2.06 & \textbf{2.64}\\
    \midrule
    \multicolumn{1}{c|}{\multirow{4}[1]{*}{\textbf{DOUBAN}}}& \multicolumn{1}{c|}{Recall@5} &28.92 &44.25 &52.66 &55.53 &56.79 &57.96 &59.43 &59.46 &62.08 &62.22 &65.81 & \underline{65.85} &\underline{65.85} & \textbf{67.37}\\
    &  \multicolumn{1}{c|}{Recall@10} &30.10 &46.98 &56.88 &57.42 &58.55 &58.21 &59.92 &59.87 &62.44 &62.52 &66.02 & 66.10 &\underline{66.11} & \textbf{67.62}\\
    &  \multicolumn{1}{c|}{MRR@5} &25.44 &36.85 &49.76 &55.12 &56.31 &57.35 &58.84 &58.70 &60.14 &60.08 &62.91 & 62.99 &\underline{63.04} & \textbf{64.68}\\
    &  \multicolumn{1}{c|}{MRR@10} &29.62 &39.57 &51.60 &56.00 &57.47 &57.71 &59.08 &59.01 &61.00 &60.77 &63.08 & \underline{63.21} &63.12 & \textbf{64.77}\\
    \midrule
    \bottomrule
    \end{tabular}
    }
  \label{tab:resutls}
\end{table*}

\subsection{Overall Performance (RQ2)} \label{subsec:overallresults}
\noindent To validate the recommendation performance of our lightweight proposal, we compared EA-GPS with thirteen state-of-the-art baseline methods on five real-world datasets, and the experimental results are reported in Table \ref{tab:resutls}. Then, we have the following observations:

1) The performance of sequential recommendation methods (i.e., GRU4REC, NAIS, SASRec, and BERT4Rec) is comprehensively superior to that of traditional recommendation methods (i.e., BPR-MF, and NCF), demonstrating the necessity of modeling user historical behavior sequences. 
2) The graph-based sequential recommendation methods (i.e., NGCF, LightGCN, SGL, TGT, SURGE, PTGCN, and GCL4SR) consistently outperform conventional sequential recommendation methods (i.e., GRU4REC, NAIS, SASRec, and BERT4Rec). This observation demonstrates the significance of considering the complicated associations among items in non-Euclidean space for SR scenarios. 
3) The TGT model outperforms most other GSRs (except EA-GPS) across a majority of evaluation metrics, indicating that the Graph Transformer is effective in capturing global dependencies within graphs. EA-GPS achieves superior performance compared to the TGT, suggesting that our proposed lightweight solution is comparable to the Graph Transformer while modeling global dependencies among nodes.
4) EA-GPS achieves the best performance in each evaluation metric on all datasets. This observation indicates that the lightweight strategy employed in EA-GPS does not compromise the prediction accuracy; instead, it enhances the performance of GSRs.
5) EA-GPS significantly outperforms all the \ac{GSRs}, highlighting the effectiveness of the external encoder in enhancing global associations for node representation learning. Moreover, it verifies the effectiveness of the positional prompt-based decoder in capturing user-specific sequential preferences from the abstract node representation space.
6) EA-GPS exhibits a better performance than TGT and PTGCN, which are two GSR methods with indirect positional encoding while learning the sequence representations. This observation indicates the superiority of our proposed positional prompt-based decoder in leveraging positional information of items within sequences.
\subsection{Ablation Studies (RQ3)}\label{subsec:ablation}
\noindent To explore the importance of different components in EA-GPS, we conduct a series of ablation experiments on three datasets (i.e., BOOK, MOVIELENS, and DOUBAN) while keeping all the hyper-parameter fixed. Specifically, we compare EA-GPS with its following variants: 
\begin{itemize}
    \item{$\text{GPS}_{OPT}$:} This is a variant method of EA-GPS that removes the positional prompt-based decoder. Instead, it just leverages a simple max pooling layer to decode the node representations.
    \item{$\text{GPS}_{RPE}$:} This variant treats the relative positional encoding (i.e., the positional encoding in Transformer) as prompts and also utilizes the masked attention network to learn the sequence representations.
    \item{$\text{GPS}_{OMA}$:} This is another variant method of EA-GPS that disables the masked attention component and directly add the positional information to the item embeddings.
    \item{$\text{GPS}_{OEA}$:} This variant method removes the external encoder while learning node representations for users and items.
    \item{$\text{GPS}_{SA}$:} This is a variant of EA-GPS that replaces the external attention component with Self-Attention (SA) mechanism.
    \item{$\text{GPS}_{LA}$:} Compare to $\text{GPS}_{SA}$, this variant replaces the SA with the Linear Attention (LA) mechanism \cite{ChoromanskiLCZS22}. Both SA and LA have been introduced in Sec~\ref{subsubsec:external}. 
    \item{$\text{GPS}_{Basic}$:} This variant of EA-GPS disables both the external encoder and the prompt-based decoder. It is basically equivalent to the LightGCN with additional consideration for sequential relationships between items.
\end{itemize}
\begin{table*}[t]
  \centering
  \normalsize
   \caption{The experimental results $(\%)$ of ablation studies on three real-world datasets.}
   \resizebox{\linewidth}{!}{ 
    \begin{tabular}{lcccc|cccc|cccc}
    \toprule
    \midrule
    \multicolumn{1}{c}{\textbf{Dataset}} & \multicolumn{4}{c|}{\textbf{BOOK}} &
    \multicolumn{4}{c|}{\textbf{MOVIELENS}} &
    \multicolumn{4}{c}{\textbf{DOUBAN}} \\
    \midrule
    \multicolumn{1}{c}{\multirow{2}[1]{*}{\textbf{Metric}}}& \multicolumn{2}{c}{\textbf{Recall}} & \multicolumn{2}{c|}{\textbf{MRR}}& \multicolumn{2}{c}{\textbf{Recall}} & \multicolumn{2}{c|}{\textbf{MRR}} & \multicolumn{2}{c}{\textbf{Recall}} & \multicolumn{2}{c}{\textbf{MRR}} \\
    \cmidrule{2-13}
    & \textbf{@5} & \textbf{@10} & \textbf{@5} & \textbf{@10}
    & \textbf{@5} & \textbf{@10} & \textbf{@5} & \textbf{@10}
    & \textbf{@5} & \textbf{@10} & \textbf{@5} & \textbf{@10}
          \\
    \midrule
    $\text{GPS}_{OPT}$ &36.45 &37.23 &31.06 &31.18
    &3.94 &7.16 &1.88 &2.05 &61.85 &61.99 &60.43 &60.81 \\
    $\text{GPS}_{RPE}$ &37.55 &38.20 &32.20 &32.34 &4.32 &7.56 &2.08 &2.43 &66.10 &66.28 &62.59 &62.89\\
    $\text{GPS}_{OMA}$ &36.80 &37.40 &31.09 &31.22
    &4.06 &7.25 &1.91 &2.08
    &62.55 &62.86 &60.95 &61.24 \\
    $\text{GPS}_{OEA}$  &36.90 &37.42 &31.18 &31.27
    &4.14 &7.30 &1.94 &2.16 
    &63.25 &63.84 &61.58 &61.84 \\
    $\text{GPS}_{SA}$ &37.28 &37.81 &31.77 &32.25
    &4.22 &7.41 &2.01 &2.33
    &64.84 &65.31 &62.12 &62.44 \\
    $\text{GPS}_{LA}$ &37.25 &37.59 &31.68 &32.05
    &4.20 &7.37 &1.99 &2.28
    &64.79 &64.99 &62.04 &62.17 \\
    $\text{GPS}_{Basic}$ &36.02 &36.91 &30.87 &31.13
    &2.88 &4.88 &1.72 &1.92 &59.43 &59.92 &59.08 &58.84 \\
    \midrule
    \textbf{EA-GPS} &\textbf{38.11} &\textbf{38.77} &\textbf{32.59} &\textbf{32.74}
    &\textbf{4.47}&\textbf{7.87}&\textbf{2.19}&\textbf{2.64}
    &\textbf{67.37}&\textbf{67.62}&\textbf{64.68}&\textbf{64.77}\\
    \midrule
    \bottomrule
    \end{tabular}
    }
  \label{tab:ablation_studies}
\end{table*}

The experimental results on three datasets are reported in Table \ref{tab:ablation_studies}. Then we have the following observations: 
1) EA-GPS outperforms $\text{GPS}_{OEA}$ and $\text{GPS}_{Basic}$. It reveals the importance of incorporating a global weighting strategy in node representation learning and highlights the effectiveness of the external attentive graph encoder in EA-GPS. 
2) EA-GPS performs better than $\text{GPS}_{OPT}$ and $\text{GPS}_{OMA}$. This observation proves the positive impact of the positional prompt-based decoder on learning user-specific sequence-level representations. EA-GPS outperforms $\text{GPS}_{OMA}$, demonstrating the effectiveness of the masked attention network within the decoder. 
3) EA-GPS outperforms the $\text{GPS}_{RPE}$, demonstrating that the absolute positions of interactions in the sequences have a stronger indicative effect than the relative positional encoding. The positional prompt-based encoder in EA-GPS performs better in capturing the evolution of user preferences. 
4) $\text{GPS}_{OEA}$ performs better than $\text{GPS}_{OPT}$. This observation suggests that the positional prompt-based decoder plays a more significant role throughout the EA-GPS model, with its importance even surpassing that of the external encoder. This finding further supports our view that the positional information of items within sequences can further unearth the potentials of the sequence recommender.
5) The performance of EA-GPS surpasses that of $\text{GPS}_{LA}$ and $\text{GPS}_{SA}$, indicating that the external encoder excels in capturing global correlations compared to LA and SA. Furthermore, this result also suggests that our proposed positional prompt-based decoder is more effective on GCNs equipped with a external encoder (i.e., EA-GPS) than on those with a primary graph encoder utilizing other global weighting strategies (i.e., self-attention or linear attention mechanism).

\begin{figure}[t]
    \centering
    \includegraphics[width=1\columnwidth]{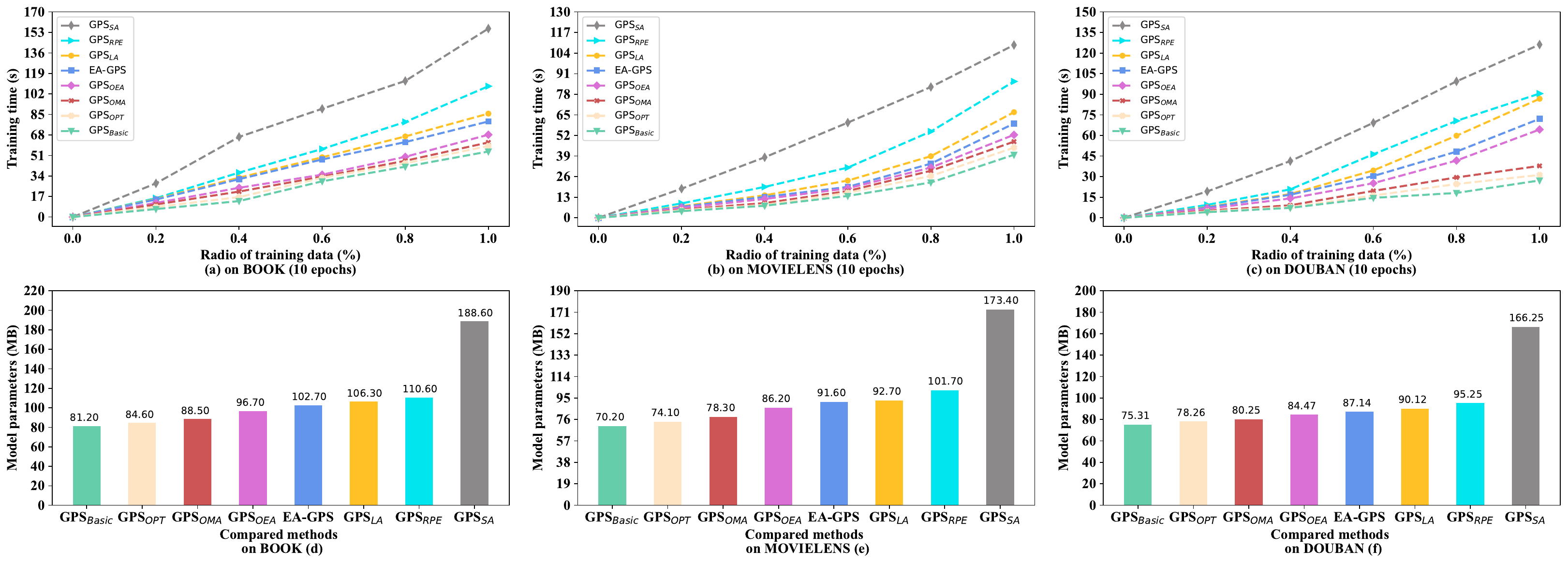}
    \caption{Time consumption and the parameter scale of EA-GPS and its variants.}
    \label{fig:ablation_LA}
    \Description{Time consumption and the parameter scale of EA-GPS and its variants.}
\end{figure}

Moreover, we adopt the same experimental strategies outlined in Section \ref{subsec:training} to assess the lightweight performance (i.e., the training time consumption and parameter scale) of EA-GPS and its variants. The experimental results on BOOK, MOVIELENS, and DOUBAN are presented in Fig. \ref{fig:ablation_LA}. Then, we have the following observations:
1) As shown in Fig. \ref{fig:ablation_LA} (a) to (c), both EA-GPS and $\text{GPS}_{OPT}$ have lower training time consumption than $\text{GPS}_{SA}$ and $\text{GPS}_{LA}$. This observation indicates that our external attention-based graph encoder has better training efficiency compared to the self-attention-based or linear attention-based methods. 
2) We also observe that EA-GPS and the variants of its key components (i.e., $\text{GPS}_{OPT}$, $\text{GPS}_{OMA}$, and $\text{GPS}_{OEA}$) exhibit more stable growth trends of the time consumption than the other variant methods (i.e., $\text{GPS}_{Basic}$, $\text{GPS}_{LA}$ and $\text{GPS}_{SA}$). This observation indicates that the core components in EA-GPS are more scalable when dealing with large-scale datasets.
3) EA-GPS requires less training time than $\text{GPS}_{RPE}$, demonstrating that the sequence decoder based on the absolute positional prompts has better training efficiency than that based on the relative positional encoding.
4) As shown in Fig. \ref{fig:ablation_LA} (d) to (f), it can be observed that the EA-GPS and $\text{GPS}_{LA}$ have smaller parameter scales than $\text{GPS}_{SA}$, and the EA-GPS performs relatively better. This is mainly because the overhead of maintaining two external memory units is significantly lower than generating key and value vectors for each input features.
5) We also observe that there is no significant difference in the parameter scale between EA-GPS and its key components (i.e., $\text{GPS}_{OPT}$, $\text{GPS}_{OMA}$, and $\text{GPS}_{OEA}$). This experimental result indicates that the parameter scale of each core component is small, providing a positive response to the \textbf{RQ3}.

\begin{figure}[t]
    \centering
    \begin{minipage}[t]{1\linewidth}
        \centering
        \includegraphics[width=0.95\columnwidth]{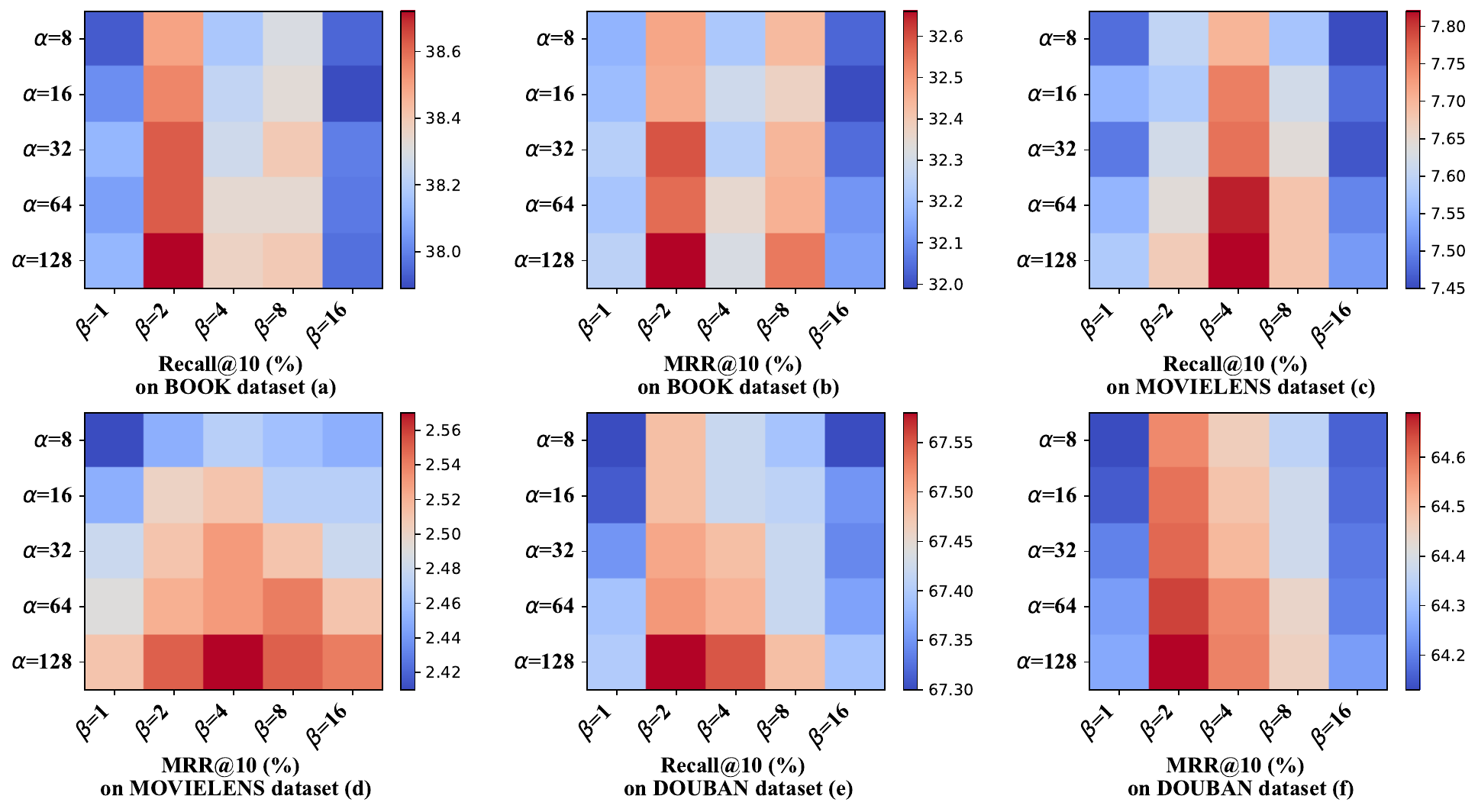}
    \end{minipage}
    \begin{minipage}[t]{1\linewidth}
        \centering
        \includegraphics[width=0.95\columnwidth]{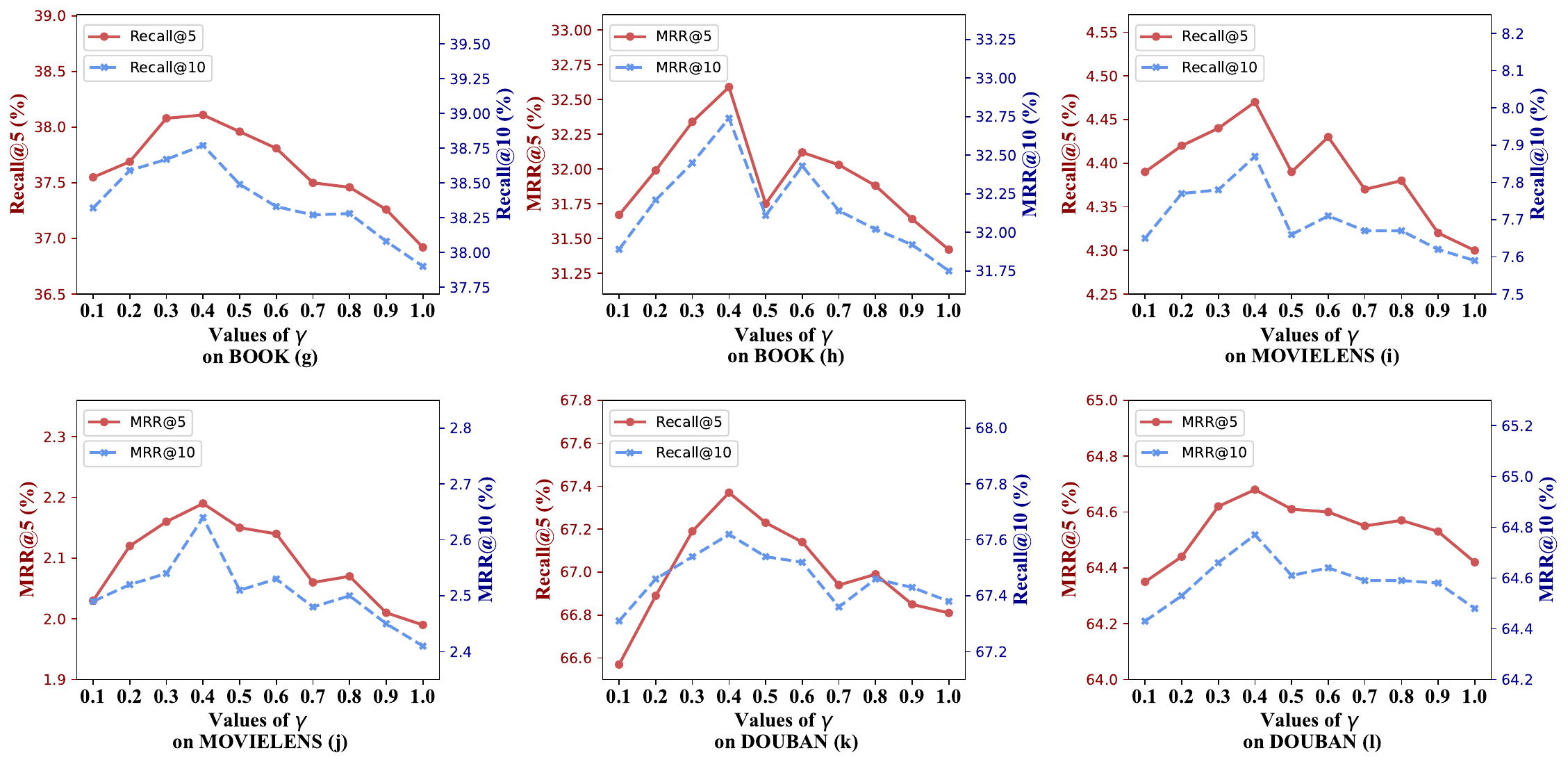}
    \end{minipage}
    \caption{Impact of hyper-parameters $\alpha$, $\beta$ and $\gamma$ on three datasets.}
    \label{fig:hyper_param}
    \Description{Impact of hyper-parameters $\alpha$, $\beta$ and $\gamma$ on three datasets.}
\end{figure}

\subsection{Impact of Three Hyper-parameters (RQ4)\label{sec:hyper_param}}
\noindent In this section, three hyper-parameters (i.e., $\alpha$, $\beta$ and $\gamma$) are introduced, which significantly impact the key components of EA-GPS (i.e., the external encoder and the positional prompt-based decoder). We conduct a series of experiments to analyze the performance variations of these hyper-parameters on the BOOK, MOVIELENS, and DOUBAN datasets.

\textbf{Impact of $\alpha$ and $\beta$.} Notably, $\alpha$ and $\beta$ are two important hyper-parameters, which control the dimension of the external memory units and the head number of the multi-head external attention mechanism, respectively. As shown in Fig. \ref{fig:hyper_param} (a) to (f), the experimental results on the heat maps demonstrate that the \acf{EA}-based global weighting strategy for enhancing node representation learning is contingent on selecting the proper value for $\alpha$ and $\beta$. The detailed observations are summarized as follows. 
1) It is evident that the better the performance of EA-GPS, the larger the value of $\alpha$. However, we also found that even when $\alpha$ takes a smaller value (such as 8 or 16), it can still achieve acceptable performance (i.e., still higher than the sub-optimal GSR baseline TGT). Moreover, the performance shift between different values of $\alpha$ is not significant, indicating that the performance of EA-GPS is not sensitive to the dimension size of the external memory unit. 
2) When $\alpha$ is set to 16, the computational complexity of external attention component is equal to the linear attention mechanism. But the performance of EA-GPS in such settings is better than $\text{GPS}_{LA}$. This observation further proves that the external attention-based encoder excels in capturing global correlations.
3) $\beta=2$ is optimal for the BOOK and DOUBAN datasets, while $\beta=4$ yields the best results for the MOVIELENS dataset. The primary factor contributing to this observation may be the varying levels of user diversity in interests across different datasets. The greater the number of attention heads required, the higher the average level of user diversity in interests. For instance, user interests in the MOVIELENS dataset are more complex and varied compared to those in the BOOK or DOUBAN datasets, which aligns with objective norms.

\begin{figure}[t]
    \centering
    \includegraphics[width=0.95\columnwidth]{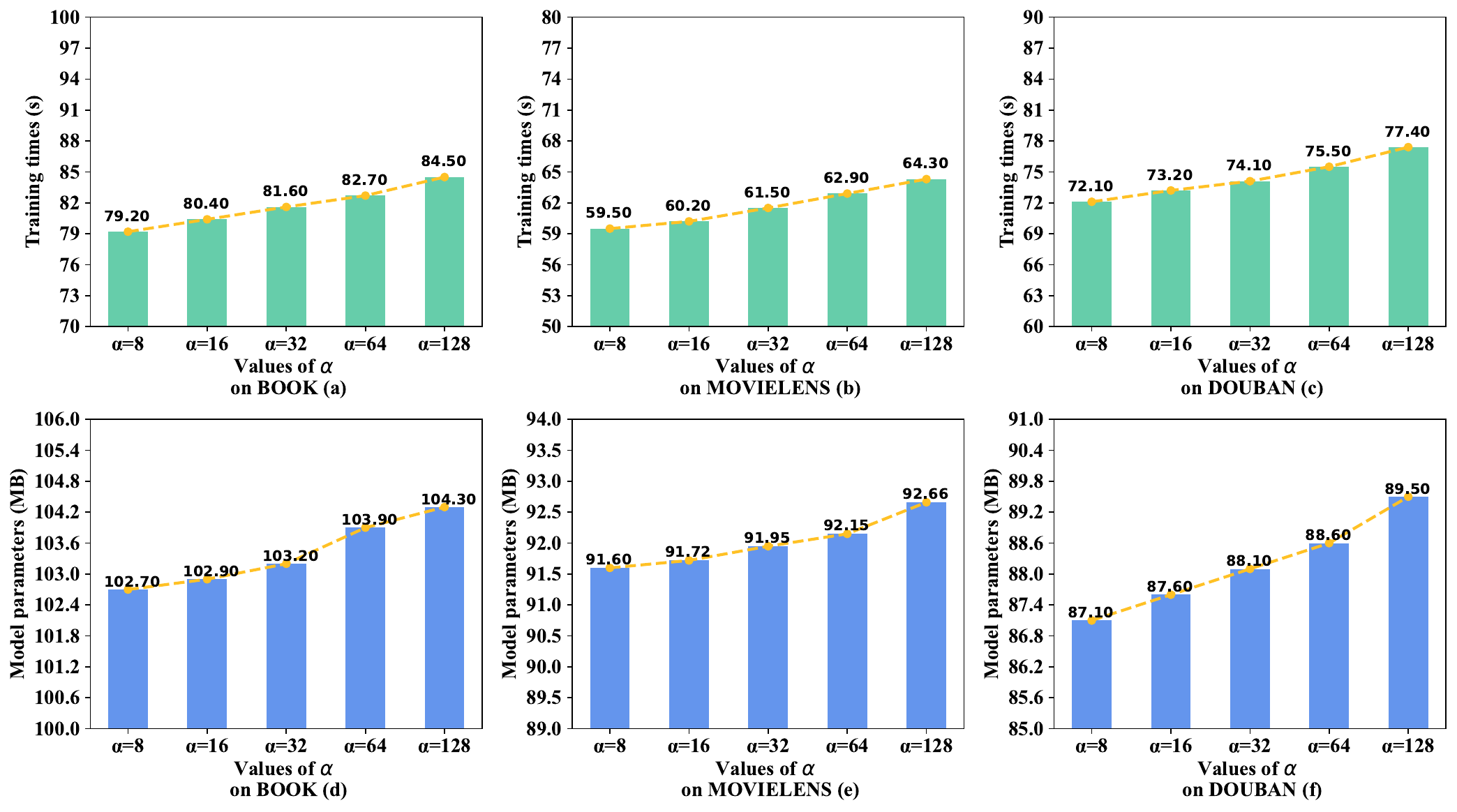}
    \caption{Time consumption and the parameter scale of EA-GPS with different value of $\alpha$.}
    \label{fig:hyper_light}
    \Description{Time consumption and the parameter scale of EA-GPS with different value of $\alpha$.}
\end{figure}

Additionally, we evaluate the impact of the hyper-parameter $\alpha$ on the lightweight performance, with the experimental results presented in Fig. \ref{fig:hyper_light}. The experimental results on these datasets indicate that a multiplicative increase in $\alpha$ does not significantly affect the training efficiency and parameter scale of EA-GPS. This observation further validates the superiority of the external encoder in providing lightweight sequential recommendations.

\textbf{Impact of $\bm{\gamma}$.} $\gamma$ is another significant hyper-parameter that controls the proportion of masked items during the sequential masking. From Fig. \ref{fig:hyper_param} (g) to (l), we discovered that the model achieves optimal performance when $\gamma$ is set to 0.4 across all three datasets, and the trends in the four evaluation metrics exhibit similar patterns on each dataset. This observation indicates that masking the prompt templates at a ratio of 0.4 is a near-universal setting. Setting it too high (e.g., $\gamma=1$) or too low (e.g., $\gamma=0.1$) results in sub-optimal performance, indicating that only an appropriate level of masking can effectively leverage the positional prompts to enhance the sequence representation learning.

\section{Conclusions and Future Work}
\noindent In summary, this paper introduces a lightweight yet efficient graph-based sequential recommendation approach named EA-GPS. We propose a lightweight external encoder that works in parallel with primary graph encoder to efficiently capture the global associations among items. We also present an efficient positional prompt-based decoder to capture the complicated positional dependencies from sequences. 
Experimental results on five real-world datasets demonstrate that EA-GPS has a smaller parameter scale and less computational complexity while significantly outperforming other state-of-the-art baseline methods.

One further exploration of this study is to conduct online testing in real-world scenarios. In the future, we plan to deploy EA-GPS into actual running online recommendation systems (e.g., shopping apps, digital TV platforms, etc.). This integration will enable us to assess the practical utility of EA-GPS. Moreover, since the primary goal of EA-GPS is to achieve a lightweight recommendation, we intend to test the model on resource-constrained edge devices, such as mobile terminals with limited memory or online servers with limited bandwidth. The performance in real-world scenarios will guide the future refinement of EA-GPS, thereby making it better serve the sequential recommendation scenarios.

\begin{acks}
\noindent This research is supported by the National Natural Science Foundation of China (Grant No. 62472263, 62072288), the Taishan Scholar Program of Shandong Province, Shandong Youth Innovation Team, the Natural Science Foundation of Shandong Province (Grant No. ZR2024MF034, ZR2022MF268).
\end{acks}

\bibliographystyle{ACM-Reference-Format}
\bibliography{main}

\end{document}